\documentclass[5p]{elsarticle}

\usepackage{xcolor}
\usepackage{listings}
\usepackage{comment}
\usepackage{amsmath}
\usepackage{amssymb}
\usepackage{algorithmic}

\usepackage{caption}
\usepackage{subcaption}
\usepackage{multirow}
\usepackage{url}
\usepackage{array}
\usepackage[toc,page]{appendix}

\newcolumntype{C}[1]{>{\centering\arraybackslash}m{#1}}

\usepackage[colorinlistoftodos,textsize=tiny,textwidth=12mm]{todonotes}

\newcommand{\fakepar}[1]{\medskip\noindent\textit{#1~~}}

\usepackage{titlesec}

\titleformat{\subsubsection}[runin]
{\normalfont\itshape}
{}{0em}{}[ -- ~~]

\journal{Computer Networks}

\begin{document}

\begin{frontmatter}

\title{Performance Benchmarking of State-of-the-Art Software Switches for NFV}

\author[1]{Tianzhu~Zhang\corref{cor1}\fnref{fn1}}
\ead{tianzhu.zhang@telecom-paristech.fr}
\author[1]{Leonardo Linguaglossa} \ead{linguaglossa@telecom-paris.fr}
\author[2]{Paolo Giaccone} \ead{paolo.giaccone@polito.it}
\author[1]{Luigi Iannone} \ead{luigi.iannone@telecom-paris.fr}
\author[1]{James Roberts} \ead{james.walter.roberts@gmail.com}

\cortext[cor1]{Corresponding author}
\fntext[fn1]{Tianzhu Zhang was affiliated with Telecom Paris during the time of writing.} 
\address[1]{Telecom Paris, 19 Place Marguerite Perey, 91120 Palaiseau, Paris, France}
\address[2]{Politecnico di Torino, Corso Duca degli Abruzzi, 24, 10129 Torino TO, Italy}

\begin{abstract}
With the ultimate goal of replacing proprietary hardware appliances with Virtual Network Functions (VNFs) implemented in software, Network Function Virtualization (NFV) has gained popularity in the past few years.
Software switches are widely employed to route traffic between VNFs and physical Network Interface Cards (NICs). It is thus of paramount importance to compare the performance of different switch designs and architectures. 
In this paper, we propose a methodology to compare fairly and comprehensively the performance of software switches. We first explore the design spaces of 7 state-of-the-art software switches and then compare their performance under four representative test scenarios. Each scenario corresponds to a specific case of routing NFV traffic between NICs and/or VNFs. In our experiments, we evaluate the throughput and latency between VNFs in two of the most popular virtualization environments, namely virtual machines (VMs) and containers. Our experimental results show that no single software switch prevails in all scenarios. It is, therefore, crucial to choose the most suitable solution for the given use case. At the same time, the presented results and analysis provide a more in-depth insight into the design tradeoffs and identify potential performance bottlenecks that could inspire new designs.
\end{abstract}

\begin{keyword}
Network Function Virtualization (NFV), Virtual Network Functions (VNF), Service Function Chain (SFC), software virtual switch, performance benchmarking methodology, high-speed packet I/O.
\end{keyword}

\end{frontmatter}

\sloppy


\section{Introduction}
\label{sec:introduction}


For many years developers have used software packet processing for fast prototyping and functional testing but have relied on the superior performance of proprietary hardware for product deployment. The limitations of commercial off-the-shelf (COTS) servers, whose general-purpose kernels and chips were not optimized for packet processing, outweighed the flexibility advantage of software solutions. This situation has changed in recent years, thanks mainly to the impulsion of Software-Defined Networking (SDN) and Network Function Virtualization (NFV) but also due to advances in the performance of COTS hardware.
It is now widely accepted that significant savings on both CapEx and OpEx can be realized on replacing expensive, proprietary, and inflexible hardware middleboxes by software counterparts.

%

A major spur to progress has been the development of high-speed I/O frameworks (e.g., DPDK~\cite{dpdk}, PF\_RING ZC~\cite{pfring}, and net\-map~\cite{rizzo2012netmap}) that employ acceleration techniques, like kernel-bypassing, polling, buffer pre-allocation, and batch processing, to achieve performance comparable to that of proprietary hardware appliances. Furthermore, modern COTS servers are equipped with multiple cores to promote parallelization and Non-Uniform Memory Access (NUMA) awareness to enhance memory access efficiency.
Software switches have largely benefited from the combined use of these acceleration techniques. They are thus widely employed by NFV platforms as their data plane to flexibly steer traffic through various components.
For example, 
Metron~\cite{katsikas2018metron}, SplitBox~\cite{asghar2016splitbox}, and MiddleClick~\cite{barbette2018building} incorporate FastClick~\cite{barbette2015fast} for traffic steering. 
UNO~\cite{le2017uno} and eVNF~\cite{van2020accelerating} leverage OVS-DPDK~\cite{ovs-dpdk} for efficient packet forwarding.  
E2~\cite{palkar2015e2} and ParaBox~\cite{zhang2017parabox} adopt BESS~\cite{han2015softnic} as data plane, while
ClickOS~\cite{martins2014clickos} and HyperNF~\cite{yasukata2017hypernf} opt for the VALE switch~\cite{rizzo2012vale}.
NetBricks~\cite{panda2016netbricks} even integrates both BESS and OVS-DPDK to benefit from their respective features.

While interest in software switches is soaring, the relative merits of different proposals are still not well-understood in the absence of comprehensive, comparative performance analysis. It is indeed a daunting task to perform such an evaluation~\cite{fang2018evaluating}, and most published comparisons relate to a small number of switch proposals~\cite{emmerich2018throughput,shanmugalingam2016dpdk} or execute a limited number of test scenarios~\cite{paolino2015snabbswitch}. Our work's objective is to propose a methodology for comparing switch performance in terms of essential metrics like throughput and latency. Our methodology aims to provide a fair comparison of a broad range of state-of-the-art software switches in a set of representative yet straightforward test scenarios. Such a methodology must take into account the different design choices that guided the switch designs. For instance, Open vSwitch (OVS)~\cite{ovs} was tailored to support match/action semantics;
VPP was constructed as a full-fledged software router;
while other solutions such as Snabb, FastClick, and BESS embraced modular design to compose complex network services.

The contribution of our work is summarized as follows: 
\begin{itemize}
\item We propose an experimental methodology to thoroughly evaluate the performance of software switches in the context of NFV.
\item We apply our methodology to compare the performance of 7 state-of-the-art software switches, namely Open vSwitch DPDK (OVS-DPDK)~\cite{ovs-dpdk}, FastClick~\cite{barbette2015fast}, Berkeley Extensible Software Switch (BESS)~\cite{han2015softnic}, VALE~\cite{rizzo2012vale,honda2015mswitch}, Snabb~\cite{paolino2015snabbswitch}, t4p4s~\cite{laki2016high,voros2018t4p4s}, and FD.io VPP~\cite{vpp}.
\item All the experiments are conducted in the same environment with identical infrastructural settings to guarantee a fair comparison. 
\end{itemize}

We start by analyzing the design space of these frameworks to build a basic understanding of their respective designs. We then define four test scenarios, as introduced in~\cite{zhang2019benchmarking,zhang2019comparing}, to provide meaningful results for different segments of a Service Function Chain (SFC).
Finally, we provide experimental measurements of throughput and latency. 
In particular, the throughput is measured with both unidirectional and bidirectional traffic. It is important to note that these experimental results depend significantly on the testbed's particular hardware configurations and software versions.
For instance, VPP achieved varied throughput on different COTS servers~\cite{csit,zhang2019comparing}. Similar phenomena were also reported in \cite{kawashima2017evaluation,zhang2019flowatcher}.
Indeed, the experimental results differ even after an OS upgrade on the same server.
Therefore, our aim is not to assess the performance in absolute terms for the adopted testing platform, but rather to define a proper comparison methodology and to identify possible performance impairments with approximate results when the switches are deployed for traffic routing in an NFV environment.

Traditionally, VNFs are usually deployed in virtual machines (VMs) based on hypervisors such as QEMU/KVM~\cite{hwang2015netvm} and Xen~\cite{martins2014clickos}. Recently, there has been an increasing trend to distribute VNFs into lightweight containers~\cite{zhang2016opennetvm,hong2017considerations,sun2017nfp}.
Our experiments consider both virtualization techniques. In particular, we choose the QEMU/KVM hypervisor to instantiate VMs and the Docker engine to manage containers. Since the Linux kernel stack imposes non-negligible overhead for both VM and container networking~\cite{rizzo2013speeding,suo2018analysis}, we further incorporate a set of VNF I/O techniques (e.g., vhost-user backend, virtio-user frontend, netmap's ptnet/veth) for different virtualized environments. 
To facilitate reproducibility, all the scripts and instructions of our experiments have been released on GitHub~\cite{software-switches}. We strongly encourage researchers and developers to use them to repeat the same set of experiments on their servers and to build on this basis to gain further understanding.

The paper is organized as follows. In Sec.~\ref{sec:related}, we review related literature on software switches and their performance comparison. Then, in Sec.~\ref{sec:space}, we explore the design space of the 7 considered software switches and highlight their specificities. In Sec.~\ref{sec:testmet}, we describe the proposed performance benchmarking methodology and the related testbed settings.
Experimental results are presented and discussed in Sec.~\ref{experiments}. Finally, we draw conclusions in Sec.~\ref{conclusion}. We also provide \ref{sec:config} to clarify a few essential settings for the considered switches.


\section{Related work}
\label{sec:related}

In this section, we survey the panoply of open-source software switches and discuss  related work on the performance comparison of different implementations.

\subsection{Software switches}
\label{subsec:rw:switches}
We first introduce the seven software switches that are directly compared in our work. Then we briefly describe some software switches that are excluded from our performance comparison study and explain the particular reason for exclusion.
%


\subsubsection{Evaluated Software Switches}
Evaluated Software Switches –The state-of-the-art soft-ware switches included in our study have been chosen for the availability of an up-to-date codebase and their promised high performance.

\noindent \textbf{OVS-DPDK}~\cite{ovs-dpdk} is a high-speed variant of OVS. Just like OVS, it supports standard management interfaces and programmable control of traffic forwarding. 
By migrating the OVS fast path into user space and adopting DPDK poll-mode drivers for packet retrieval, it completely avoids the overhead imposed by the general-purpose kernel stack and interrupt-based packet receiving.

\noindent \textbf{t4p4s}~\cite{laki2016high,voros2018t4p4s} is a platform-independent software switch specifically designed to support P4 semantics~\cite{bosshart2014p4}. It consists of a compiler and a hardware abstraction layer. 
The former generates optimized switching code from P4 programs, while the latter deals with platform-specific function deployment and optimization. Currently, t4p4s is capable of generating switching programs on multiple hardware targets, including Intel X86 with the DPDK library.

\noindent \textbf{FastClick}~\cite{barbette2015fast} extends the codebase of Click Modular Router~\cite{kohler2000click} and integrates packet I/O techniques such as DPDK and netmap for high-speed processing.
Its data path is also optimized leveraging various acceleration techniques, including zero-copy, batching, and multi-queueing.

\noindent \textbf{Snabb}~\cite{paolino2015snabbswitch} is a high-speed modular software switch with a collection of predefined modules enabling the composition of advanced network applications. Like MoonRoute~\cite{gallenmuller2017building}, it is based on Lua and LuaJIT~\cite{luajit}. Snabb is known for introducing the vhost-user protocol~\cite{vhost-user}, which features direct packet exchange between user-space software switches and VNFs running on VMs, without kernel intervention.

\noindent \textbf{BESS}~\cite{han2015softnic} is a modular software switch from UC Berkeley featuring a set of built-in modules used to compose network services. Modules can be glued together and fed to the daemon process, which deals with packet scheduling (enabling traffic prioritization) and processing. 


\noindent \textbf{VPP}~\cite{vpp} is a software router that allows users to configure the forwarding graph and process packets in batches. It incorporates many throughput optimization techniques while also supporting interrupt mode when using native drivers.
Besides, VPP provides a CLI for experimentation and debugging while resorting to the binary API for production use (bindings to C, C++, Go, Python over a non-blocking, shared memory interface).


\noindent \textbf{VALE}~\cite{rizzo2012vale} is an L2 software switch based on netmap. It adopts batch computing and memory prefetching to enhance processing efficiency. mSwitch~\cite{honda2015mswitch} augments VALE with enhanced switching logic.  A pass-through approach named \texttt{ptnet} was later proposed to ensure high-speed packet delivery between virtual machines~\cite{maffione2016flexible}.
In contrast to most of the other switches that employ poll-mode drivers and complete kernel bypass, VALE presents better compatibility with OS kernels and relies on system calls and interrupts for packets' I/O operations. Therefore, it is interesting to compare VALE with other solutions.

\subsubsection{Other Software Switches}\label{sec:irr}
In this part, we briefly reference some software switches that are excluded from our comparison study for different reasons.
\textbf{RouteBricks}~\cite{dobrescu2009routebricks} achieves multi-Gigabit/s packet processing speeds by exploiting parallelization both within and across commodity servers.
\textbf{Pa\-cket\-Sha\-der}~\cite{han2011packetshader} boosts packet processing using graphics processing units (GPUs).
\textbf{Hyper-Switch}~\cite{ram2013hyper} improves packet forwarding between virtual machines and Xen hypervisor by adopting batch processing and computation offloading. 
\textbf{Cuckoo Switch}~\cite{zhou2013scalable}, a software Ethernet switch, adopts the cuckoo hashing algorithm for forwarding table lookup and DPDK for packet I/O operations, thus realizing both memory efficiency and high-speed processing.
{\bf IVS~\cite{ivs}} is an OpenFlow software switch based on the OVS kernel module purposed to achieve efficient pipeline processing with the Floodlight SDN controller.
\textbf{MoonRoute}~\cite{gallenmuller2017building} is a software router based on MoonGen~\cite{emmerich2015moongen} and LuaJIT~\cite{luajit}. The adoption of Lua scripting language improves programmability compared to other software switches using low-level languages such as C or C++.
Despite their interesting features, we exclude these switches from the direct quantitative comparison due to their outdated codebase.

\textbf{VFP}~\cite{firestone2017vfp} is designed to support SDN in cloud data centers.
\textbf{PVPP}~\cite{choi2017pvpp} extends VPP to support P4 programs. {\bf ESwitch}~\cite{molnar2016dataplane} employs template-based code generation to optimize the OpenFlow pipeline. 
{\bf VOSYSwitch}~\cite{fanguede2018novel} extends Snabb to support OpenFlow semantics and achieves performance comparable to OVS-DPDK.
{\bf NF-switch}~\cite{hsieh2017nf} embeds a many-field flow table inside OVS to implement efficient SFCs.
{\bf OVS eBPF}~\cite{tu2017building} is a kernel-based datapath optimization for OVS with extended Berkeley Packet Filter (eBPF).
The work in~\cite{tseng2017accelerating} seeks to accelerate OVS processing with integrated GPU.
These solutions are excluded for direct comparison because of the unavailable codebase.

\textbf{PISCES}~\cite{shahbaz2016pisces} extends Open vSwitch with the support of the P4 language. \textbf{HyperV}~\cite{zhang2017hyperv} is another P4 dataplane hypervisor, and its DPDK target achieves comparable performance to PISCES.
However, as detailed in \cite{voros2018t4p4s}, t4p4s outperforms this by a factor of two when running the baseline L2 forwarding application. We thus only consider t4p4s in our comparison.
\textbf{Lagopus}~\cite{rahimi2016high} and {\bf xDPd-DPDK}~\cite{doriguzzi2014datapath} are two user-space OpenFlow switches that integrate DPDK for high-speed processing.
Due to their overlapping functionalities and lower performance with OVS-DPDK~\cite{kawashima2017evaluation}, we exclude them from our comparison to avoid duplicated effort.
In \cite{bonelli2017acceleration}, \textbf{OfSoftSwitch} was optimized with the PFQ framework~\cite{bonelli2012multi}. However, this switch also presented limited performance ($\leq$ 4~Mpps with 64B packets) and is therefore not included in our comparison. 
A similar observation also applies for {\bf BOFUSS}~\cite{fernandes2019road} and {\bf KVS}~\cite{choi2017kvs}.
BOFUSS provides an SDN software switch implementation with performance no better than standard OVS.
KVS utilizes RSS hash to index flow rules and only achieves 2.68 Gbps on a single core.

Finally, \textbf{ClickNF}~\cite{laufer2016climb,gallo2018clicknf} extends Click with a set of modules enabling complex L2 to L7 network functions. Since ClickNF is similar to FastClick in terms of design and performance, we do not consider it to avoid duplicates.
{\bf vNS}~\cite{Gallo:2018:VMP:3234200.3234242} implements a full-fledged modular network stack based on ClickNF to provide tailored functionalities for network service composition. 
\textbf{REdge}~\cite{hayakawa2019resilient} extends VALE to build high-performance NFV backend. 
{\bf Oko~\cite{chaignon2018oko}} extends OVS-DPDK to integrate runtime stateful filtering and monitoring eBPF programs into the OpenFlow processing pipeline.
Although these proposals have implemented many interesting features on the original switches, they do not bring any performance enhancement in the baseline test scenarios defined by our comparison study.

\subsection{Performance Comparison}
\label{subsec:rw:performance}

\begin{table*}[!tp]
\caption{Performance comparison of existing software switches under the NFV environment. We highlight in bold the software switches relevant to our comparison study, as explained in Sec. \ref{sec:irr}. {\em ``Bare-metal"} refers to the case of configuring software switches to forward packets between two physical NICs, {\em ``Inter-VNF forwarding"} refers to traffic forwarding between two VNFs. {\em ``SFC"} refers to the service function chaining scenario with a varied number of sequential VNFs, {\em ``Uni."} refers to unidirectional throughput test, {\em ``Bi."} refers to bidirectional throughput test. $\qquad\qquad$ $^{(\star)}$Only OVS (and not OVS-DPDK) was evaluated with containerized VNFs in~\cite{bonafiglia2015assessing}.}
\vspace{-0.5cm}
\scriptsize
\begin{center}
\setlength\tabcolsep{1pt}
\begin{tabular}{|c|c||c|cc|c|c||cc|c|c|}
\hline
\multirow{2}{*}{Ref.} & \multirow{2}{*}{Software switches under test} & \multirow{2}{*}{Bare-metal} & \multicolumn{2}{c|}{VNF environment} & Inter-VNF & \multirow{2}{*}{SFC} & \multicolumn{2}{c|}{Throughput} & \multirow{2}{*}{Latency} \\ 
&&  & VM & Container & forwarding & & Uni. & Bi. & \\ \hline \hline
\cite{hong2017considerations} & \bf BESS & & & \checkmark & & \checkmark & \checkmark && \checkmark \\ \hline
\cite{bonafiglia2015assessing} & OVS, \bf OVS-DPDK & \checkmark & \checkmark & \checkmark$^{(\star)}$ & & \checkmark & \checkmark && \checkmark \\ \hline
\cite{casoni2013performance} & Linux bridge, \bf VALE & & & \checkmark &  \checkmark & \checkmark  & \checkmark & & \\ \hline
\cite{niu2017unveiling} & ClickOS, \bf BESS & & & & & \checkmark & \checkmark && \checkmark \\ \hline
\cite{emmerich2018throughput} & Linux bridge, OVS, \bf OVS-DPDK & \checkmark & \checkmark & & \checkmark & & \checkmark & & \checkmark \\ \hline
\cite{voros2018t4p4s} & OVS, PISCES, \bf t4p4s & \checkmark& &&&& \checkmark && \\ \hline
\cite{pitaev2017multi,pitaev2018characterizing} & \bf OVS-DPDK, VPP & & \checkmark & &  & & \checkmark & & \\ \hline
\cite{fang2018evaluating} & \bf BESS, VPP, OVS-DPDK & \checkmark & & & & & \checkmark && \\ \hline
\cite{lettieri2017survey} & \bf OVS-DPDK, Snabb, VALE & & \checkmark & & \checkmark & & \checkmark &  & \\ \hline
\cite{paolino2015snabbswitch} & {\bf Snabb}, OVS, {\bf OVS-DPDK}, Linux bridge &  & \checkmark & & \checkmark & & \checkmark && \\ \hline
\cite{kawashima2016host,kawashima2017evaluation} & {\bf OVS-DPDK}, OVS, Linux bridge, Lagopus, xDPd-DPDK & \checkmark & \checkmark & & &  & \checkmark && \checkmark \\ \hline \hline
CSIT-1904 & \bf VPP & \checkmark & \checkmark & \checkmark & \checkmark & \checkmark & \checkmark & \checkmark & \checkmark \\ \hline
VSperf & OVS, \bf OVS-DPDK, VPP & \checkmark & \checkmark & \checkmark &  \checkmark & \checkmark & \checkmark & \checkmark & \checkmark \\ \hline \hline
\bf Our work & \bf OVS-DPDK, FastClick, Snabb, VPP, t4p4s, BESS, VALE & \checkmark & \checkmark & \checkmark & \checkmark & \checkmark & \checkmark & \checkmark & \checkmark \\ \hline
\end{tabular}
\end{center}
\label{tab:comp}
\end{table*}%

The literature includes several works aiming to evaluate the performance of software switches.
Rojas et al.~\cite{rojas2012evaluation} measured the throughput of several customized software routers on two workstations equipped with 10/100 Mbps NICs.
Emmerich et al.~\cite{emmerich2014performance} compared OVS throughput with Linux bridge and Linux kernel IP forwarding. According to their results, the standard OVS failed to attain 2~Mpps with 64B packets.
Shanmugalingam et al.~\cite{shanmugalingam2016dpdk} evaluated the throughput of OVS-DPDK with port mirroring using 1 Gbps NICs.
Our work differs in that we only focus on software switch implementations capable of achieving much better performance (e.g., more than two orders of magnitude higher throughput).

Some prior performance comparison works are particularly relevant to ours.
Fang et al.~\cite{fang2018evaluating} analyzed the bare-metal throughput of BESS, VPP, and OVS-DPDK when forwarding traffic between two physical interfaces.
Pitaev et al.~\cite{pitaev2017multi,pitaev2018characterizing} compared the throughput of VPP and OVS-DPDK with heterogeneous VNFs in VMs. 
Lettieri et al.~\cite{lettieri2017survey} compared the throughput and CPU utilization of VALE, OVS-DPDK, and Snabb. They configured the switches to steer traffic between two VNFs and between a VNF and a physical NIC. We do not characterize CPU utilization because VALE operating in hybrid mode (polling and interrupt) always outperforms other switches running in polling mode.
In addition, since our work focuses solely on software switches, we attach physical NICs to the VALE switch, not directly to the VMs. 
Paolino et al.~\cite{paolino2015snabbswitch} compared the throughput of Snabb, Open vSwitch, OVS-DPDK, and Linux bridge under the same test scenarios as \cite{lettieri2017survey}.
Note that all the previous works did not compare software switches in terms of the processing latency, which is another critical performance metric, especially for realtime services.
Emmerich et al.~\cite{emmerich2018throughput} evaluated both throughput and latency of OVS and OVS-DPDK forwarding packets between two physical NICs, between two VNFs in VMs, and between a VNF and a physical NIC.
Kawashima et al.~\cite{kawashima2017evaluation,kawashima2016host} evaluated the throughput and latency of OVS, Linux Bridge, OVS-DPDK, Lagopus, and xDPd-DPDK forwarding synthetic traffic between two physical NICs and between a VNF and a physical NIC.
These works did not compare performance in the presence of service function chains (SFCs) with multiple VNFs.

Casoni et al. \cite{casoni2013performance} compared the throughput for VALE and Linux bridges forwarding packets between a set of sequential/parallel LXC containers.
Bonafiglia et al.~\cite{bonafiglia2015assessing} evaluated throughput and latency for OVS and OVS-DPDK on SFCs of varied lengths. However, they still relied on the inefficient kernel path for containers. Our experiments adopt more efficient I/O techniques for containerized VNFs, as explained in Sec. \ref{sub:virtualizationtech}.
Hong et al.~\cite{hong2017considerations} measured the throughput and latency of BESS for service function chaining. They wrapped Click programs inside containers and deployed them as VNFs. As indicated by their results, the Click programs imposed a non-negligible processing overhead that outweighs the impact the BESS. Our work instead uses more lightweight VNFs to better reflect the forwarding capacity of software switches.
Niu~et~al.~\cite{niu2017unveiling} compared the throughput and latency of BESS and ClickOS in SFC scenarios. We preferred to consider VALE rather than ClickOS in our comparison, as the latter is a full-fledged NFV framework rather than a software switch. Furthermore, in our comparison, all VMs were based on the KVM hypervisor, avoiding the uncertainty arising when one system uses KVM and the other Xen.
Besides, none of the aforementioned works considered the throughput test with bidirectional traffic, which, based on private communication with Cisco engineers, is another critical factor. Therefore, we evaluate the throughput of software switches with both unidirectional and bidirectional traffic.
In general, our work provides a more comprehensive performance comparison for software switches in NFV.
As highlighted in Table~\ref{tab:comp}, our work considers a more complete set of state-of-the-art software switches and compares them in more test scenarios with different virtualization techniques and performance metrics.

In contrast to the existing literature, in addition to providing measurement results, our work seeks to define a comparison methodology. Such a methodology consists of test scenarios and metrics designed to enable a deeper understanding of software switch performance and help identify potential bottlenecks. 
There are two open-source projects, namely FD.io CSIT-1904~\cite{csit-1904} and VSperf~\cite{vsperf}, that are very relevant to our work.
CSIT-1904 aims to define a comprehensive set of test scenarios for VPP and DPDK applications.
VSperf, proposed by the Open Platform for NFV Project (OPNFV), focuses on the benchmarking methodology of virtual switches for the NFV infrastructure~\cite{rfc8204}. Currently, it has integrated vanilla OVS, OVS-DPDK, and VPP.
Our work covers all the test scenarios defined by the two projects.
Moreover, the reported experimental results relate to a set of representative, state-of-the-art software switches that are more extensive than any performance comparison studies in prior work.



\section{Software switch design space}
\label{sec:space}

\begin{table*}[!t]
  \caption{Taxonomy of state-of-the-art high-performance software switches}
  \vspace*{-0.1cm}
  \label{tab:taxonomy}
  \centering
  \scriptsize
  \setlength\tabcolsep{3pt}
  \begin{tabular}{|c|cc|c|cc|c|c|c|c|}
\hline
 & \multicolumn{2}{c|}{\bf Architecture} &  \bf Programming & \multicolumn{2}{c|}{ \bf  Model} & \bf Virtual & \bf Runtime & \bf Programming & \bf Main\\
 &  Self-contained        &  Modular        &  \bf Paradigm  &RTC & Pipeline  & \bf Interface & \bf Reprogrammability & \bf Language & \bf Purpose\\
\hline
\hline
BESS &           &    \checkmark      		&   Structured   	& \checkmark	& \checkmark    & vhost-user & High & C, Python & Programmable NIC \\
\hline

Snabb &           &   \checkmark       		&   Structured & 	&  \checkmark & vhost-user & High & Lua, C & VM-to-VM\\
\hline

OVS-DPDK & \checkmark          &          	&  Match/action   & \checkmark	&   & vhost-user & High & C & SDN switch \\
\hline

FastClick &           &   \checkmark       &  Structured    & \checkmark	&  & vhost-user & Medium & C++ & Modular router \\
\hline

VPP &     \checkmark      &          		&  Structured  & \checkmark	&   &vhost-user & Medium & C & Full router \\
\hline

VALE &     \checkmark      &         &  Structured  & \checkmark	&   & ptnet & Low & C & Virtual L2 Ethernet\\
 \hline

t4p4s & \checkmark          &          	&  Match/action   & \checkmark	&   & vhost-user & Low & C, Python & P4 switch \\
\hline
\end{tabular}

\end{table*}

We first discuss the importance of exploring the different design objectives of alternative software switches before considering how the seven representative state-of-the-art solutions fit into a design space taxonomy.

\subsection{Design Objectives}
Before performing a comparative evaluation, it is very important to understand the main design differences between the considered software switches.
This may require identifying the adopted processing model, or ascertaining whether the switch has been designed for a particular application such as SDN or NFV. Such a task is time-consuming but appears an essential precondition to avoiding biased results or an incorrect interpretation of the impact of subtle, performance impacting details.

Rather than providing a detailed discussion of implementation and/or acceleration techniques, for which we refer to the survey in~\cite{linguaglossa2019survey}, we aim in this section to consider each switch design in relation to a number of technical aspects affecting packet processing performance. The objective is to gain insight on how to devise meaningful experimental scenarios. A summary of this taxonomy is shown in Table~\ref{tab:taxonomy}, whose details are now discussed.

\subsection{Architecture}
\label{subsec:sp:arch}

A significant difference between software switches derives from the way packet processing is configured and, more importantly, executed.
A {\em self-contained} architecture is defined as a full-fledged software that can be deployed with minimal configuration effort.
The switch data path is predefined, though modifications at compile time are allowed, and all processing functions are deployed in a single process.
In contrast, a {\em modular} architecture targets a high degree of flexibility. This is usually achieved by providing a set of predefined, well-known network functions that can be arranged in a {forwarding graph}. The latter
can even be re-configured at runtime when each node is a different thread or process or extended with custom network functions.

Our evaluation takes into account four switches designed with a self-contained architecture: VALE~\cite{rizzo2012vale}, VPP~\cite{vppcommag2018}, t4p4s~\cite{laki2016high}, and OVS-DPDK~\cite{ovs-dpdk}.
VALE is an L2 learning switch based on netmap, which can interconnect both physical NICs and virtual interfaces and forward packets at high speed. Though it is feasible to connect VALE with an external program,
it is considered here as a self-contained architecture.
VPP consists of a forwarding graph with hundreds of functions and support for additional plugins~\cite{vppcomnet2019}.
It exposes a command-line interface that can be used to configure the router with a syntax similar to the Cisco IOS operating system.
OVS-DPDK is a software switch built for SDN in which packet processing is realized via a set of match/action tables (cf.\ Sec.~\ref{sub:paradigm}), which can be modified via the
\texttt{ovs-vsctl} API.
Custom packet processing can be realized by adding new code that must be compiled inside the original codebase.
t4p4s is designed to support P4~\cite{bosshart2014p4} semantics, whose  workflow is quite similar to OVS-DPDK. It consists of a parsing stage on packet entry and a de-parsing stage when packets exit. Match/action tables, described through P4, are deployed between these two stages to indicate the sequence of operations to perform on packets.

The other switch designs considered in our study, FastClick \cite{barbette2015fast}, BESS~\cite{han2015softnic}, and Snabb~\cite{paolino2015snabbswitch}, belong to the modular category.
FastClick, one of the latest versions of the original Click Modular Router, consists of a set of nodes that can be arranged using a Click-specific configuration language.
BESS also has a modular architecture, although the modules are more general and less specialized than those of FastClick.
Similarly, Snabb interconnects modules with links to compose network services.

\subsection{Design Paradigm}
\label{sub:paradigm}
Software switch implementations are heavily influenced by their target use cases.
We classify the design paradigm into two categories.
The first one adopts structured programming to route traffic across VNFs, as done by a majority of existing software switches.
The second solution is to use the match/action programming paradigm exploited by OVS-DPDK and t4p4s.
Packet processing is realized using built-in packet classification algorithms that match specific header fields and apply the corresponding actions.

\subsection{Processing Model}
\label{sub:rtcvspipeline}

When packets are delivered to a software switch, there are generally two ways to process them: \textit{run-to-completion} (RTC) and \textit{pipeline}.
The former refers to a model in which a single thread performs full packet processing before being forwarded or discarded, while the latter refers to a model according to which packets go through several threads, each containing a portion of processing logic, to complete full processing.

Most frameworks (VPP, OVS-DPDK, t4p4s, and VALE) adopt the run-to-com\-ple\-tion model to reduce the context switching overhead.
Even FastClick, an extension of Click designed with a pipeline model in mind, has completely moved to a full run-to-completion approach.
Snabb is the only considered switch that processes packets solely according to a pipeline model, while BESS can adopt either model (RTC by default) depending on the implemented multicore approach. 

\subsection{Virtual Interfaces}
\label{sub:virtualizationtech}

Software switches mainly rely on virtual interfaces to interact with VNFs and realize the intended traffic steering on NFV platforms. There are several techniques for VM and container networking.
Most of the VMs based on QEMU/KVM communicate with the outside world using the virtio~\cite{russell2008virtio} standard. It consists of the {\tt virtio\_net} para-virtualized frontend network driver and the {\tt vhost\_net} backend driver.
Traditionally, vhost\_net takes packets into the kernel and copies them back to the user-space software switch. However, this is not desirable from a performance point of view. To address this issue, Snabb implements {\tt vhost-user}, a backend driver allowing direct packet exchange between user-space software switches and VMs. Compared with vhost\_net, vhost-user provides better performance as it eliminates the overhead imposed by the kernel. DPDK also adopts this solution and hence all of the frameworks considered in this work, except VALE that is based on netmap, use \texttt{vhost-user}~\cite{vhost-user} as backend driver\footnote{Note that ivshmem was an alternative VM networking for DPDK applications. We do not consider it in our study as it has been removed in newer versions of DPDK.}.
VALE relies on \texttt{ptnet} for efficient VM networking~\cite{maffione2016flexible}. \texttt{ptnet} is a new para-virtualized device driver that grants the VMs direct access to packet buffers of netmap ports on the host using the netmap API. Compared with vhost-user, ptnet delivers packets in a zero-copy manner without incurring the overhead of queueing (as for virtio) or packet descriptor format conversion, at the cost of a lower degree of host-VM isolation and more difficult live migration.

Compared to virtual machines that emulate resources at the hardware level, the container is an alternative lightweight solution at OS level and achieves isolation through namespaces/cgroups.
For high-speed container networking, the DPDK community advocates using the {\tt virtio-user} frontend driver~\cite{virtio-user}. {\tt virtio-user} is a shared memory mechanism based on virtio. It implements a vhost adapter to emulate virtio ports and bridge vhost backend drivers, without the involvement of any hypervisor. As demonstrated by \cite{tan2017virtio}, {\tt virtio-user} manages to achieve more than $3.5\times$ performance boost over the standard kernel-based approaches. We thus consider {\tt virtio-user} as the virtual device driver for all the software switches using vhost-user backend.
Netmap instead provides native support for the veth interface~\cite{veth}. By specifying the kernel source path during compilation, the veth.ko module based on netmap optimization is automatically created.
Container networking in netmap mode is made possible by moving {\tt veth} pairs across different Linux namespaces.
In our tests, we attach one end of veth to a VALE switch and move the other end to the namespace of a Docker container, to implement peer-to-peer zero-copy packet delivery. To guarantee optimal performance, both sides of a veth pair must be attached to netmap applications.

\subsection{Runtime Reprogrammability}
\label{sub:reprogrammability}

Although software switches are usually easy to program, it is also important to consider their degree of \textit{reprogrammability}.
As an example, programmable packet processors can be written as a simple C program. However, adding a new feature may require rewriting part of the code and, sometimes, to also rerun, recompile or replace binary executables.
However, a highly reprogrammable software switch should offer the possibility to change the internal processing pipeline at runtime with no need for recompilation.
We categorize the software switches into three degrees of reprogrammability: high, medium, and low.

Snabb, BESS, and OVS-DPDK have the highest degree of reprogrammability. Thanks to the App engine and command-line tools,
Snabb can interactively load standard modules to adjust its processing pipeline at runtime.
BESS's {\tt bessctl} utility serves the same purpose. 
The behavior of OVS-DPDK can also be modified at runtime. In particular, external SDN controllers can populate flow rules to the OVS match/action tables through southbound protocols such as OpenFlow~\cite{mckeown2008openflow}.
Both VPP and FastClick allow to program some modules and execute custom packet processing applications.
In particular, the VPP command-line interface allows existing modules to be configured and new plugins to be added at runtime. Nevertheless, changing the version of the same plugin requires restarting the software switch. Therefore, VPP has a medium degree of reprogrammability. Similarly, even though some modules can be interactively configured, a FastClick instance has to be restarted when the processing graph is changed and therefore has a medium degree of reprogrammability.
Finally, both t4p4s and VALE switch have a low degree of reprogrammability since they do not provide any means to dynamically adjust their packet processing at runtime.

\subsection{Programming Language}
\label{sub:programminglanguage}
The choice of one particular programming language over another can be dictated by performance requirements, programmability, or time-to-market considerations. Most of the software frameworks for high-speed packet processing are written in C and/or C++.
Since both languages are performant, feature-rich, and portable across different platforms, most of the software switches considered in our study implement their performance-critical components using them.
High-level programming languages, such as Python and Lua, are also adopted by some software switches.
For example, BESS additionally provides a Python API to facilitate the composition of configuration scripts, t4p4s implements its P4 compiler in Python.
Snabb is mainly based on Lua and also wraps snippets of C code using LuaJIT, which profiles and optimizes code execution at runtime~\cite{paolino2015snabbswitch}. With the relatively better programmability of Lua and dynamic optimization of LuaJIT, Snabb is expected to be an efficient solution.

\subsection{Switch Primary Purpose}
\label{sub:purpose}

Packet processing frameworks can sustain good performance, thanks to a large collection of acceleration techniques discussed in the survey~\cite{linguaglossa2019survey}. The adoption of these techniques depends on the primary purpose for which the software switch has been designed. Considering this purpose is of interest for two main reasons: (i) it may provide hints on the performance of each design in some specific scenarios; (ii) it may be helpful in understanding which of the software switch implementations is more suitable for some particular user requirements.

{BESS}
provides a native way to easily schedule packets without only using the simple FIFO approach, thus enabling custom policies, resource sharing, and traffic shaping.
Resource sharing mechanisms may also be implemented on top of existing frameworks, e.g., the authors of~\cite{addanki2019controlling} implemented fair sharing of both CPU and bandwidth using fair packet dropping on top of VPP. However, to the best of our knowledge, BESS is the only design that natively provides scheduling capabilities without the need to write a custom algorithm.
{Snabb}
targets a performant and straightforward packet processing framework. Its core optimizations leverage runtime profiling and rely on LuaJIT to optimize the most frequently executed portion of code, rather than relying on the static compilation. Its app engine can dynamically register new apps, making it one of the most flexible solutions for high-speed packet processing.
Unlike other switches, it implements its own compact kernel bypass mechanism without relying on DPDK or netmap.
{OVS-DPDK} aims to provide the benefits of SDN (i.e., separation of data and control planes) with the flexibility of a software solution. Its data path is highly optimized thanks to the presence of internal flow caches.
It can also be used as a static switch with predefined rules, or as a fully functional SDN switch in conjunction with an external control plane.
%
{t4p4s} implements a high-speed, platform-independent P4 switch. Its compiler synthesizes P4 programs and generates core switch code, which is then converted to platform-specific instructions by its hardware abstraction layer. It is representative of several efforts to implement production-ready P4 switches.
{FastClick} aims to provide a high-speed modular router that can process millions of packets per second by arranging custom functions in a graph-like fashion. The advantage of FastClick is the possibility to re-arrange its rich set of internal elements to realize different types of packet processing applications.
{VPP} should be considered when a fully-featured software network function (e.g., switch, router, or security appliance) is required.
Its code was part of Cisco's high-end routers before being released as open-source and therefore contains a large set of software components that can be used for all kinds of possible L2-L4 applications.
%
%
{VALE}
fulfills the role of a high-speed L2 learning switch that interconnects  multiple VMs. Its primary purpose is to provide a high-speed virtual local Ethernet switch.


\begin{figure}[!tb]
\centering
\includegraphics[width=8.5cm]{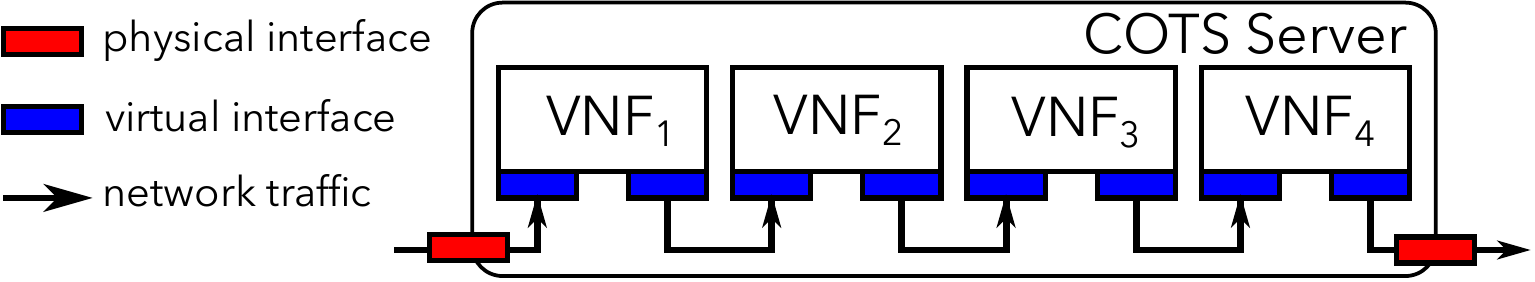}
\caption{Example of a service function chain in NFV going through different physical (p) and virtual (v) interfaces on a COTS server.}\label{fig:chain}
\end{figure}

\section{Test methodology}\label{sec:testmet}

This section shows our methodology to compare generic software switches' performance in terms of throughput and latency, i.e., two crucial metrics to evaluate the performance and scalability of NFV applications. 

When the traffic traverses a service chain of multiple VNFs, it follows a path through a sequence of interfaces which may be either physical (p) or  virtual  (v), as shown in the example of Fig.~\ref{fig:chain}.
This basic observation motivates the test scenarios considered in Sec.~\ref{sec:scenarios}, which include all possible combinations of physical and virtual interfaces.
For each scenario, we conduct an experimental measurement campaign in our platform, as described in Sec.~\ref{subsec:exp:platform}. The configuration  settings for each specific software switch considered in our work are discussed in Sec.~\ref{sec:setting}.

%
\subsection{Test Scenarios}\label{sec:scenarios}

\begin{figure}[!tb]
\begin{center}
\includegraphics[width=0.34\textwidth]{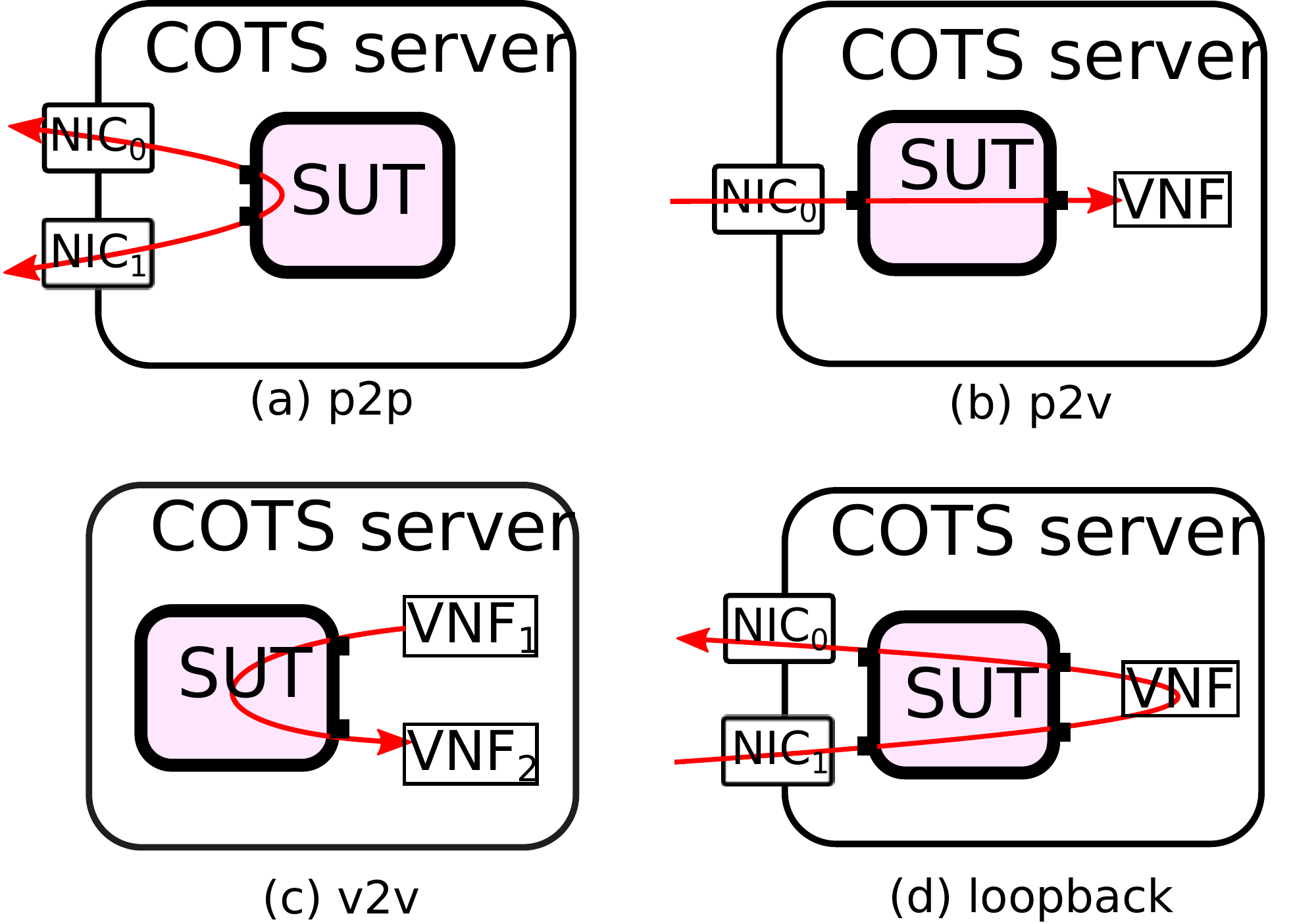}
    \caption{A logical view of the four proposed test scenarios. Red arrows illustrate the packet flow in the System Under Test (SUT), namely the software switch. }
    \label{fig:scenarios}
\end{center}
\end{figure}

For a meaningful and comprehensive comparison in a NFV system, we propose to consider four reference test scenarios, {\em p2p}, {\em p2v}, {\em v2v}, and {\em loopback}, as illustrated in Fig.~\ref{fig:scenarios}. We assume a logical server with two dual-port NICs and denote the software switch as {\bf System Under Test (SUT)}. In practice, we implement all the scenarios on a single COTS server, as described in Sec.~\ref{subsec:exp:platform}.


\fakepar{Physical-to-physical scenario (\textbf{p2p}) --}
Packets entering from one physical input interface are forwarded to the physical output interface by the software switch, as shown in Fig.~\ref{fig:scenarios}-(a). Although this scenario does not deal with VNFs, it is still relevant since common network functions are increasingly hosted by software switches, either to augment the physical NIC~\cite{han2015softnic} or to reduce duplicated VNF processing~\cite{Gallo:2018:VMP:3234200.3234242}.
Evaluating the bare forwarding rate between two physical interfaces thus provides a useful baseline reference.
Furthermore, combined with other scenarios, p2p can be used to evaluate the overhead imposed by a virtualized environment, both qualitatively and quantitatively.

\fakepar{Physical-to-virtual scenario (\textbf{p2v}) -- }
The software switch  forwards packets between a physical interface and a VNF hosted in a virtualized environment, as shown in Fig.~\ref{fig:scenarios}-(b). This scenario can be mapped to the first and last hop of VNF chains inside a server. Combined with p2p, p2v reveals the software switch performance when connected to a virtualized environment.

\fakepar{Virtual-to-virtual scenario (\textbf{v2v}) --}
The software switch transfers the traffic between two virtual interfaces, as shown in Fig.~\ref{fig:scenarios}-(c). This scenario is used to assess the traffic exchange performance by subsequent VNFs in a chain running in the same server. 
Since no physical interface is involved, the forwarding rate in this scenario is not limited by the NIC's hardware, but by the underlying bus architecture (typically front-side bus and memory bus).

\fakepar{Loopback scenario -- }
 The software switch steers incoming packets from physical NIC$_0$ through a chain of VNFs before sending them out through physical NIC$_1$. Each VNF is deployed inside a virtual machine or a container and chained with other VNFs through the virtual interfaces of the software switch.  Fig.~\ref{fig:scenarios}-(d) illustrates the case of a service chain with a single VNF. We additionally take into account service chains with multiple VNFs in our study.
 This scenario mimics a complete NFV service chain within the same server.

\medskip
When comparing the above four scenarios, it is worth noticing that the memory bandwidth bounds the throughput for the v2v scenario. In contrast, when the physical interface is involved (p2v, p2p scenarios), throughput is bounded by the NIC capacity.

%
\subsection{Measurement Testbed}
\label{subsec:exp:platform}

This section describes the hardware and software configuration that we used to implement our methodology.
This description can be used as a reference to design a measurement platform based on the available state-of-the-art technologies.

\begin{figure*}[!tb]
	\centering
	\begin{subfigure}[t]{0.23\textwidth}
		\centering
		\includegraphics[height=0.9in]{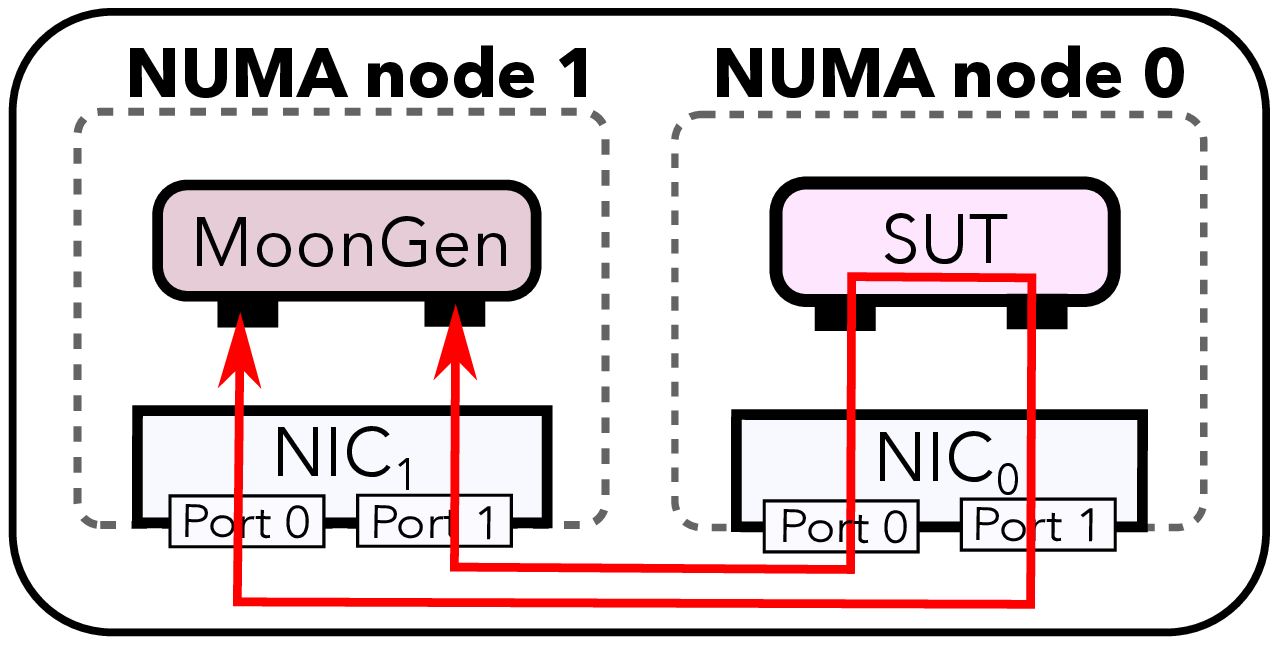}
		\caption{p2p}
		\label{fig:testbed:p2p}
	\end{subfigure}
	~
	\begin{subfigure}[t]{0.23\textwidth}
		\centering
		\includegraphics[height=0.9in]{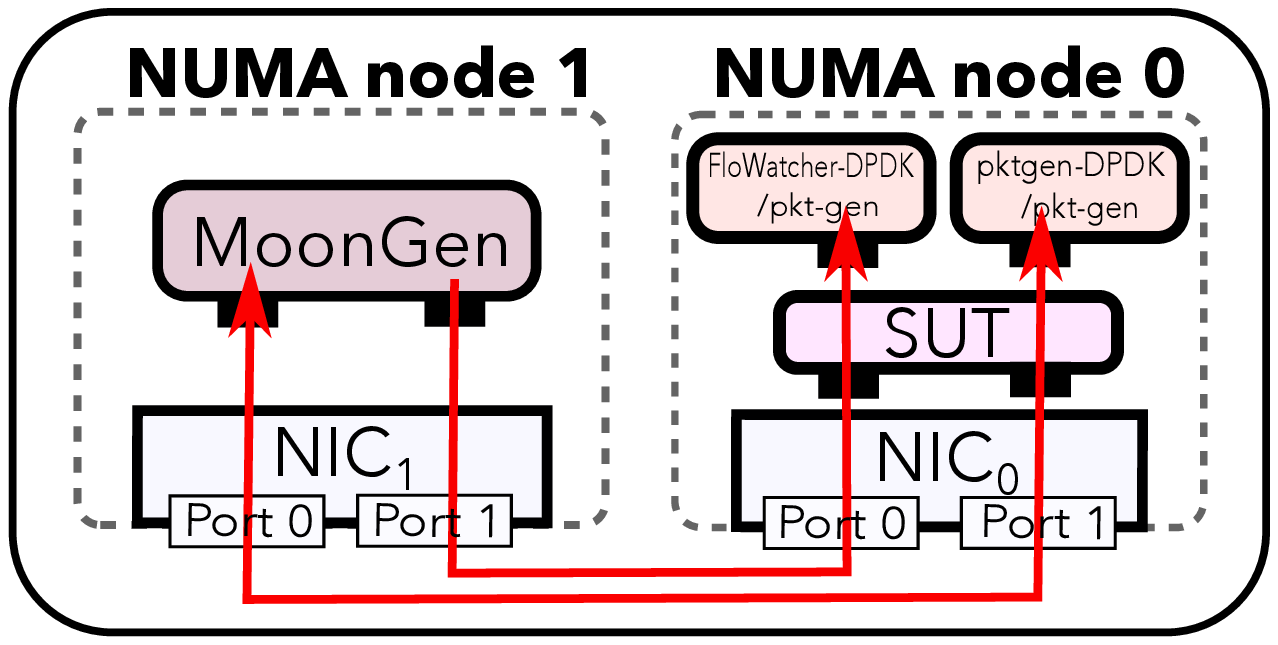}
		\caption{p2v}
		\label{fig:testbed:p2v}
	\end{subfigure}
	~
	\begin{subfigure}[t]{0.23\textwidth}
		\centering
		\includegraphics[height=0.9in]{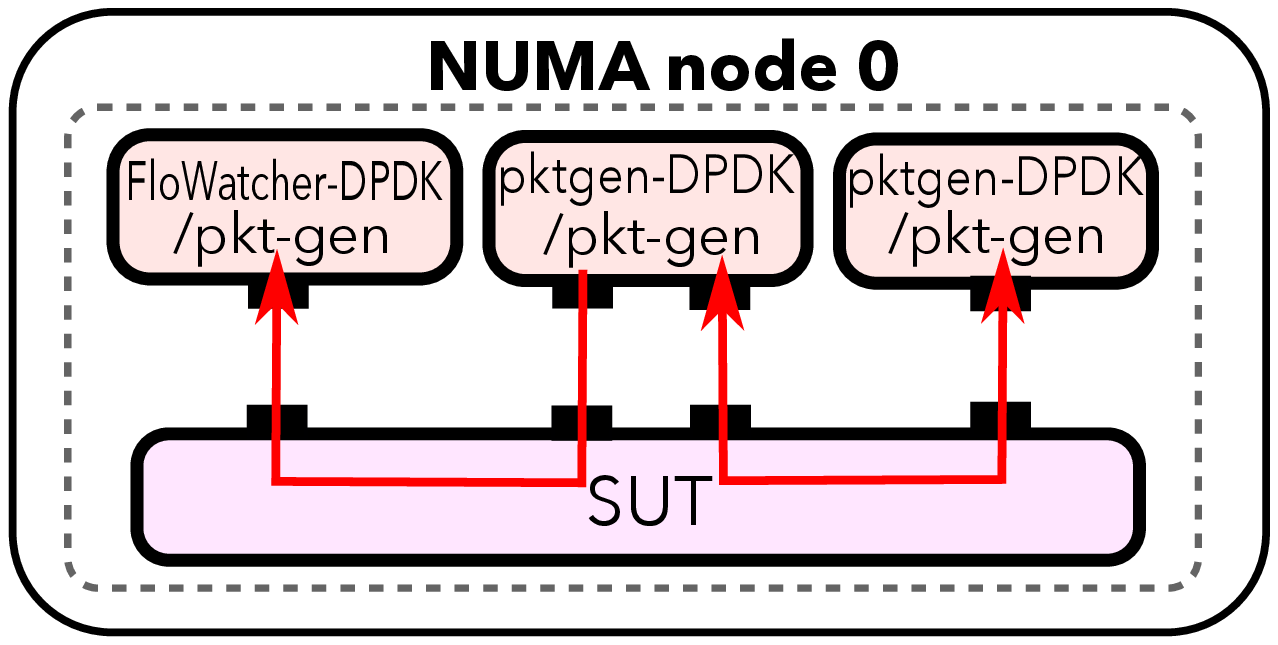}
		\caption{v2v}
		\label{fig:testbed:v2v}
	\end{subfigure}
	~
	\begin{subfigure}[t]{0.24\textwidth}
		\centering
		\includegraphics[height=0.9in]{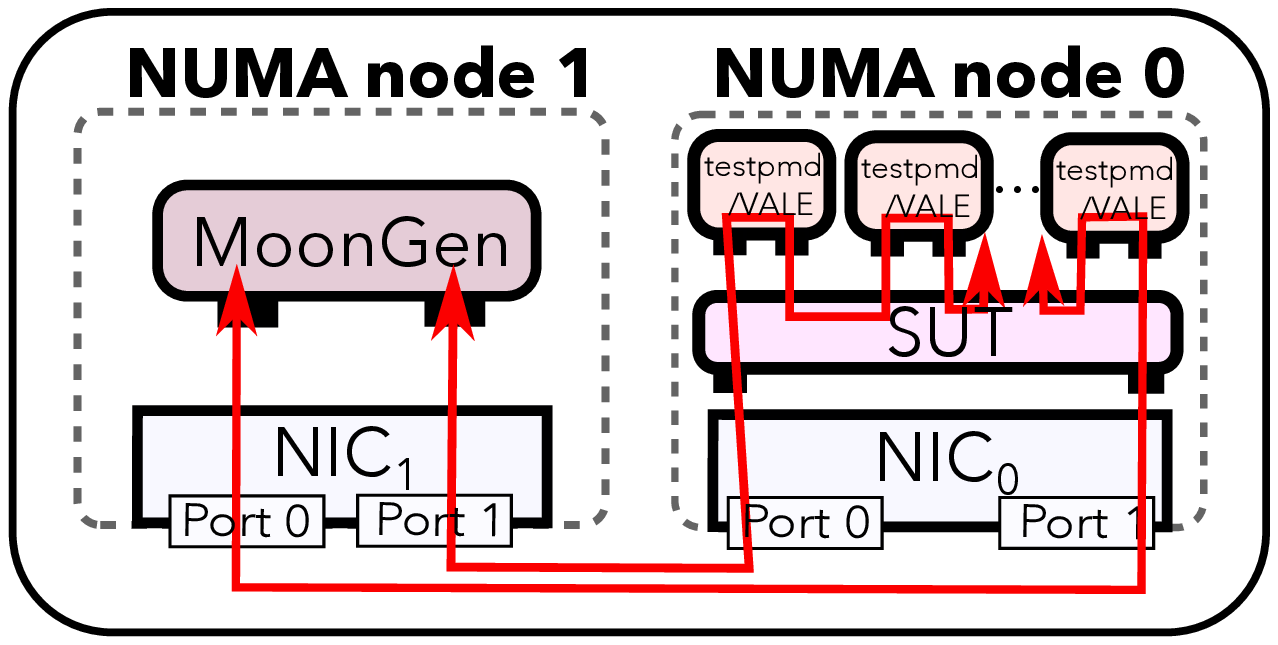}
		\caption{loopback}
		\label{fig:testbed:lo}
	\end{subfigure}
	\caption{Test scenarios mapped to our testbed with two NUMA nodes each associated with a dual-port 10~Gbps NIC directly connected to the other NUMA node's NIC. Red arrows represent the data flow. In particular, we highlight unidirectional and bidirectional traffic for p2v and v2v scenarios with two arrows respectively, as their configurations are asymmetric on TX and RX ends.}
	\label{fig:testbed}
\end{figure*}

Our testbed includes a commodity server equipped with two Intel Xeon E5-2690 v3 @ 2.60GHz CPUs (each with 24 virtual cores under hyper-threading and 32k/256k/30720K L1-3 caches), 192GB DDR4 memory @2.13GHz, and two Intel 82599ES dual-port 10Gbps NICs spread over two NUMA nodes.
The server runs Ubuntu 16.04.1 operating system with Linux 4.15.0-65-generic kernel distribution.
We deployed VNFs inside both VMs and containers to evaluate the efficiency of software switches interacting with different virtual environments.
Containers are instantiated with Docker (version~18.09.7), while VMs are launched from a CentOS~7~\cite{centos7} image using QEMU virtualizer~\cite{qemu}. In particular, we use QEMU~2.2 in experiments for BESS as newer versions present compatibility issues.
As we verified on our testbed, QEMU~2.2 yields the same throughput and latency as the newer QEMU~3.0 for the other switches.
Furthermore, as recommended in~\cite{snabb-tuning,vpp-tuning}, we fix the CPU frequency to 2.6~GHz by setting the scaling governor to ``\emph{performance}'' and disable Turboboost to reduce performance variance. We also reserve 1GB Hugepages to minimize Translation Lookaside Buffer (TLB) misses and assign 32 pages for each NUMA node.
Finally, some CPU cores are specifically isolated from the kernel scheduler using the {\tt isolcpus} boot parameter and reserved solely for the software switch under test and the VNFs.
The setup for each test scenario is illustrated in Fig.~\ref{fig:testbed}. Software switches are always deployed on a single core on NUMA node 0 to ensure a fair comparison and avoid the inter-core transfer overhead in multi-core configurations. Single core is also arguably a reasonable assumption as network operators usually seek to limit resources devoted to networking.
Each VM/container is allocated four physical cores.
We utilize a collection of high-speed software packet processing tools, including MoonGen~\cite{emmerich2015moongen}, pktgen-DPDK~\cite{pktgen-dpdk} (version~19.10.0), FloWatcher-DPDK~\cite{zhang2018flowmon}, and netmap's pkt-gen~\cite{pkt-gen}, for traffic generation and measurement. The DPDK version used for all the tests is always 18.11.3 (LTS) for both the host machine and the virtual environment.

It is important to note that the use of the same server for both traffic generation/reception and the system under test does not introduce spurious interference since the cores and memory are effectively isolated under the NUMA architecture of our server. In particular, we combine software switch utilities (e.g., handles or command-line options to tune DPDK EAL parameters) with system tools (e.g., {\tt numactl}, {\tt taskset}) to guarantee core and memory affinities.
For the v2v scenario, everything runs on NUMA node 0 without the involvement of physical NICs; thus, the traffic forwarding rate is only limited by the local memory speed.
For the other scenarios, the TX/RX components run on NUMA node 1 while the software switch under test (and TX/RX for p2v scenario) is deployed on NUMA node 0. The cores only access memory in their local NUMA node and do not share remote memory. The same benchmark-setting was also adopted in \cite{fanguede2018novel}.
Note that, since packets are transferred through physical NICs, their maximum bandwidth (10~Gbps) constitutes the theoretical bottleneck for these scenarios.
Although many works have already conducted experiments using 40/100~Gbps NICs, we argue that it is still relevant to characterize software switches in the NFV environment with 10~Gbps NICs. In fact, according to our experimental results in Sec.~\ref{experiments}, most of the software switches under test fail to sustain a 10/20~Gbps forwarding rate in most unidirectional/bidirectional test scenarios. 

%
\subsection{Software Switch Settings}\label{sec:setting}

For each tested software switch, we used the latest functional version/commit available at the time of writing, namely: FastClick (commit {\tt 9d5e9c6}); BESS (``Haswell'' archive, specifically built for Haswell CPU architecture); OVS-DPDK (version 2.11.90);  Snabb (commit {\tt 24c9a67}); VALE (netmap commit {\tt 42270fc}); t4p4s (commit {\tt b1161b2}); and VPP (version 19.04).
Moreover, as detailed in \cite{fang2018evaluating}, each software switch requires a specialized parameter setting to render the optimal performance with different input traffic and application contexts.
However, given the huge number of system and application parameters ranging from GRUB settings to internal packet buffer sizes, it is impossible to exhaustively test and find the most suitable set of values for all the test scenarios. Here we only perform basic parameter tunings for each software switch to improve performance. As the test scenarios and traffic patterns are intrinsically simple, we believe these basic tunings are sufficient to conduct a fair performance comparison study for state-of-the-art software switches. Indeed, the goal of our work is not to find the optimal throughput and processing latency in absolute terms for each considered software switch.
We also report the settings required for each scenario, taking into account the specific software switches considered in the tests. Additional configuration details are reported in \ref{sec:config}.


\subsubsection*{p2p scenario}\label{sec:p2ps}
In this scenario, the software switch acts as a packet forwarder from one physical port to the other without the involvement of any virtualization layer.
We configure MoonGen to transmit synthetic traffic at 10~Gbps from NUMA node 1 to the SUT, as illustrated in Fig.~\ref{fig:testbed:p2p}.
Similar to \cite{emmerich2018throughput,kawashima2017evaluation,hong2017considerations,bonafiglia2015assessing,lettieri2017survey}, we generate identical packets and fix the packet size to 64B, 256B, and 1024B, respectively. The results of synthetic traffic provide a fair evaluation of the sheer forwarding capability for each software switch and can also be utilized to estimate the performance of more realistic traffic patterns.
 Packets are sent at the maximum rate, disregarding any drops\footnote{Note that this is different from the usual Non-Drop-Rate (NDR) of CSIT 1904~\cite{csit}: a binary search for the NDR is not suited for evaluating software switches as it may converge to unreliable points due to even a single packet drop, caused at the driver level.}.
We consider unidirectional traffic and measure the corresponding throughput (in Gbps) on NUMA node 1, by collecting outbound traffic from NUMA node 0. We also consider bidirectional traffic, which doubles the packet processing rate that the switch has to sustain. We measure the bidirectional throughput by simultaneously transmitting packets towards both interfaces of NUMA node~0. 
We also run the baseline DPDK {\tt testpmd} sample application~\cite{testpmd} in the fastest forwarding mode (i.e., forwarding packets as they are) and measure the bare-metal throughput as a baseline reference.

\subsubsection*{p2v scenario}\label{sec:p2vs}
We configure the SUT to allow the communication between the VNF and the physical NIC, as illustrated in Fig.~\ref{fig:testbed}-(b).
Since this is a hybrid test scenario (connecting the physical and virtual environments), we consider three possible combinations of the packet workflow.
In particular, we consider a physical-to-virtual unidirectional flow (denoted as ``unidirectional''), the reverse  virtual-to-physical flow (called ``unidirectional-reverse'') and the full-duplex combination of the above cases (named ``bidirectional'').
To obtain the highest throughput for each software switch under different virtualization techniques, we apply different configurations using specific software tools and virtual device drivers.
The VALE switch requires a specific configuration, as it relies on netmap's ptnet driver for high-speed VM networking. To enable ptnet inside VMs, we use a customized QEMU provided by netmap authors~\cite{qemu-vm}, as it supports ptnet virtual interfaces with VALE ports as their host-side backend~\cite{maffione2016flexible}. We then compile netmap with ptnet support in the guest VM and run netmap applications inside the VM to maximize throughput.
Note that we are aware that ptnet also supports passthrough of physical interfaces directly, without connecting them using VALE.
However, we decided not to use this feature because our work focuses on software switches instead of device passthrough.
For high-speed networking testing between containers, netmap provides native support for {\tt veth} interfaces. We built a Docker image with netmap applications (as VNFs), for experiments related to VALE. To bridge VALE switch with the containerized VNF, we put one end of the veth pair into the container namespace and attach the other end to a VALE instance.

For both VM and container tests in the p2v scenario, we use pkt-gen as VNF inside the container. In the unidirectional test, one instance of pkt-gen in RX mode is attached to the ptnet/veth virtual interface to measure the throughput, while for the reversed test, pkt-gen runs in TX mode.
In the bidirectional test, two instances of pkt-gen in TX/RX modes are simultaneously attached to the virtual interface.
To test the throughput of other switches with VMs, we use the standard QEMU and create a virtio-pci virtual interface with a vhost-user as backend. We further accelerate packet I/O inside the guest VM by deploying DPDK and attach the virtio-pci interface to the {\tt igb\_uio} poll-mode driver.
For the case of containers, we build a Docker image wrapping the DPDK suite and run VNF on top of a virtio-user interface with vhost-user backend supplied by SUT.
We run FloWatcher-DPDK~\cite{zhang2019flowatcher}, a lightweight software traffic monitor, to measure the unidirectional throughput.
To measure the bidirectional throughput, we use pktgen-DPDK, a high-speed traffic generator/monitor. For the unidirectional-reverse test, we use pktgen-DPDK only as the traffic generation VNF.
On NUMA node 1, MoonGen is used as a traffic generator for unidirectional tests, and it is used as a traffic generator/monitor for the bidirectional tests. FloWatcher-DPDK is used to measure the reversed unidirectional throughput.
In the VM case, we configure Snabb in client mode to flexibly reconfigure Snabb without re-instantiating all the VMs. Instead, for the container case, we have to configure Snabb in server mode since virtio-user cannot create the Unix socket in the absence of the hypervisor.

To obtain bidirectional traffic, we initiate two pkt-gen instances (for VALE)/one pktgen-DPDK instance (for others) to TX/RX from inside the VM/container, and start another MoonGen instance to TX/RX simultaneously on NUMA node 1, as illustrated in Fig.~\ref{fig:testbed:p2v}.
However, we experienced severe performance degradation when the two pkt-gen instances are attached to the same ptnet port inside the VM. To overcome this, we attach the pkt-gen instances to a netmap virtual interface, which is, in turn, attached to the ptnet port through a VALE instance inside the VM. Actually, this setting imposes an extra hop of packet forwarding, but this is the best option to achieve reasonable bidirectional p2v traffic with VALE. Without this bottleneck, the real bidirectional throughput of VALE is expected to be much higher. We do not observe the same issue in the container test with two pkt-gen instances attached to the same veth interface concurrently.

\subsubsection*{v2v scenario}\label{sec:v2vs}
In the v2v scenario, as illustrated in Fig.~\ref{fig:testbed}-(c), we need to instantiate two VMs, each with a virtual interface attached to the software switch under test. The virtual interface configurations are similar to those in the p2v scenario.
We deploy a traffic generator in the first VNF and configure it to inject packets towards the software switch, which in turn forwards packets to the monitoring VNF. Similar to previous scenarios, different traffic generation/measurement tools are required to realize the intended data path for different switches.
For VALE unidirectional throughput, we deploy an instance of pkt-gen in each VM/container and configure them to perform traffic generation/measurement, respectively. For other switches, we run pktgen-DPDK in the first VM/container as a traffic generator and FloWatcher-DPDK on the second VM/container to measure unidirectional throughput.

To generate bidirectional traffic, we deploy an instance of pktgen-DPDK in each VM/container to transmit packets at maximum rate and measure the aggregated throughput, for software switches other than VALE.
For the VALE switch, we instead need two pkt-gen instances in each VM/container to transmit and receive simultaneously. Similar to the p2v bidirectional test, for VM deployment, we attach both pkt-gen instances in each VM to a netmap virtual interface, which is attached to the ptnet virtual interface through a VALE instance. For the container deployment, we directly attach both pkt-gen instances to the veth interface in each container.

\subsubsection*{Loopback scenario}\label{sec:los}
We instantiate a chain of VNFs, with  four cores and a pair of virtual interfaces allocated to each VNF.
Each software switch transfers traffic across the VNFs in sequence, forming a linear service chain. By default, a single instance of each VNF is deployed in a VM or container. Fig.~\ref{fig:testbed}-(d) illustrates the setup.
For VALE, we configure two ptnet virtual interfaces for each VM in which we run a VALE instance as a VNF. This VALE instance cross-connects the pair of ptnet ports. Each VM is linked to its successor through a VALE instance. The first and last VM also need to link the physical ports with two additional VALE instances.
Similarly, for container configuration, two pairs of veth interfaces are created, each of which has one end attached to VALE and the other end attached to a container hosting a new VALE instance as VNF. In all, we need $(N+1)$ VALE instances for a  service chain of $N$ VNFs.
For the other software switches, we configure two virtio interfaces with vhost-user backend for each VM, in which we run an instance of the DPDK testpmd sample application that cross-connects interfaces and updates the destination MAC addresses. On NUMA node~1, we start MoonGen to generate 10~Gbps traffic through one port and measure throughput for different packet sizes from the other port. For bidirectional traffic, MoonGen is configured to generate 10~Gbps traffic from both its physical ports and to measure the aggregated throughput.
The software switch is configured to transfer packets between MoonGen and the service chain. We vary the number of VNFs from 1 to 5 to test the throughput of each switch with increasing service chain length.
Regarding the switch configurations, BESS exhibits compatibility issues with QEMU 3.3.10 and cannot instantiate more than 3 VMs simultaneously. As a result, we degrade to QEMU 2.2.0 specifically for BESS in this scenario.
Furthermore, it was not possible to perform the loopback test for Snabb with containerized VNFs due to the issue reported in~\cite{snabb-question}.

\subsection{Latency Test}
\label{subsec:exp:latency1}

%
We measure the round-trip time (RTT) latency, which in our case is defined as the time spent between packet emission by the traffic generator and the time the traffic monitor receives the packet.

To avoid saturation and perform meaningful latency measurements, it is necessary to identify the Maximal Forwarding Rate ($R^+$), defined as the maximum rate the software switch can forward packets without experiencing losses. Injecting packets at a speed higher than $R^+$ causes congestion and leads to packet losses that would bias the measured latency. On the other hand, injecting packets at a very small rate may also impair latency as most solutions employ batch processing. 
It is well known that it is very hard to determine $R^+$ since software traffic generators generally lack the stability of hardware and may induce non-deterministic packet losses.\footnote{Precision is made more difficult by the coarse granularity of software traffic generators. MoonGen, for example, rounds up TX rates in the range $[9.88,10]$~Gbps to $10$~Gbps.} VNF chains in the loopback scenario tend to exacerbate this uncertainty. Rather than trying to identify the precise $R^+$, we follow the methodology introduced in~\cite{vppcomnet2019} and define $R^+$ as the average throughput achieved under saturating input conditions.
We measure latency at loads of 0.10, 0.50, and 0.99 times $R^+$. Thus, 0.99$R^+$ reflects the latency under heavy input load, 0.50$R^+$ under intermediate load, while 0.10$R^+$ shows the impact of batch processing on latency under low load.

We perform the described latency measurement specifically for p2p and loopback scenarios as, in these two scenarios, MoonGen can leverage the NIC to accurately and efficiently timestamp UDP packets~\cite{emmerich2015moongen}.
We have not performed a latency test for p2v, as its RTT is expected to be similar to that of the loopback scenario with one VNF. For v2v, it is not possible to perform the same test as for p2p and loopback since virtual interfaces, unlike physical ones, do not support hardware timestamping. Fortunately, pktgen-DPDK implements a software timestamping feature that can still be utilized in both VMs and Docker containers. Although less accurate than hardware ti\-me\-stamp\-ing, it provides a means to compare different software switches under the same setup.

\subsubsection*{p2p scenario}
To measure RTT in the p2p scenario, MoonGen is configured with two threads.
One thread generates synthetic traffic with 64B packets, as used for measuring throughput. The other TX thread periodically injects, as background traffic, Precise Time Protocol (PTP) packets with specific sequence numbers, collects these special PTP packets on their way back from the other port of the NIC in NUMA node 1, and calculates the round-trip time based on the difference between TX and RX ti\-me\-stamps. These timestamps are generated by the underlying Intel 82599 NIC, under the instruction of MoonGen.

\subsubsection*{v2v scenario}
For the v2v latency test, we cannot leverage the hardware timestamping feature of MoonGen inside virtualized environments. As a result, we have to adopt different methods for different tools to realize a relatively fair comparison.
Thanks to the good compatibility with the operating system, standard tools can be used to measure the latency for VALE in this scenario.
We simply configure routing using \texttt{ip} command for each VM. We then ping the second VM from the first and get the average RTT.
Note that we cannot do the same inside Docker containers, as the veth interfaces are always in the down state and cannot be invoked with system tools.
Other switches do not support system tools due to the complete kernel-bypassing architecture of DPDK.
Instead, as mentioned before, we measure latency using the software timestamping feature of pktgen-DPDK to measure the RTT.
The setup is the same as the bidirectional v2v throughput test: we configure one virtio-pci/virtio-user interface for each VM/container. All the interfaces are attached to the SUT.
In the first VM, we launch an instance of pktgen-DPDK with the latency test option enabled. Packets are timestamped and transmitted from one virtio interface towards the SUT, which forwards traffic to the second VM.
The second VM, in turn, bounces the packets back to the SUT using the DPDK testpmd application. Then the SUT sends the packets to the first VM.
The pktgen-DPDK instance in the first VM timestamps the received packets and calculates the RTT based on the difference between RX and TX timestamps.
We set the packet size to 96B\footnote{This is specifically required for pktgen-DPDK since smaller packet size always renders 0~$\mu$s RTT.} and transmit them at the maximal rate for all the tests.
Although not as accurate as hardware timestamping, this approach reveals the main characteristics of the solutions.

\subsubsection*{Loopback scenario}
The loopback latency test uses the same settings as the p2p test with  $R^+$ set to the corresponding unidirectional loopback throughput.
For all the switches except VALE, each VNF is essentially an instance of the DPDK testpmd application running in ``mac" forwarding mode. Similar to the throughput test, we deploy testpmd in both VMs and containers using the vhost-user backend. Again, we cannot show results for Snabb with containers here due to the issue reported in~\cite{snabb-question}.
For VALE, we again run a VALE instance as l2 forwarding VNF inside the virtualized environment.


\section{Experimental Results}
\label{experiments}

We now show the experimental results obtained using the methodology presented in Sec.~\ref{sec:testmet} to evaluate performance in terms of throughput (Sec.~\ref{sec:ttest}) and  latency (Sec.~\ref{sec:ltest}).

\subsection{Throughput Tests}\label{sec:ttest}

\begin{figure}[!tb]
  \centering
       \includegraphics[width=0.47\textwidth]{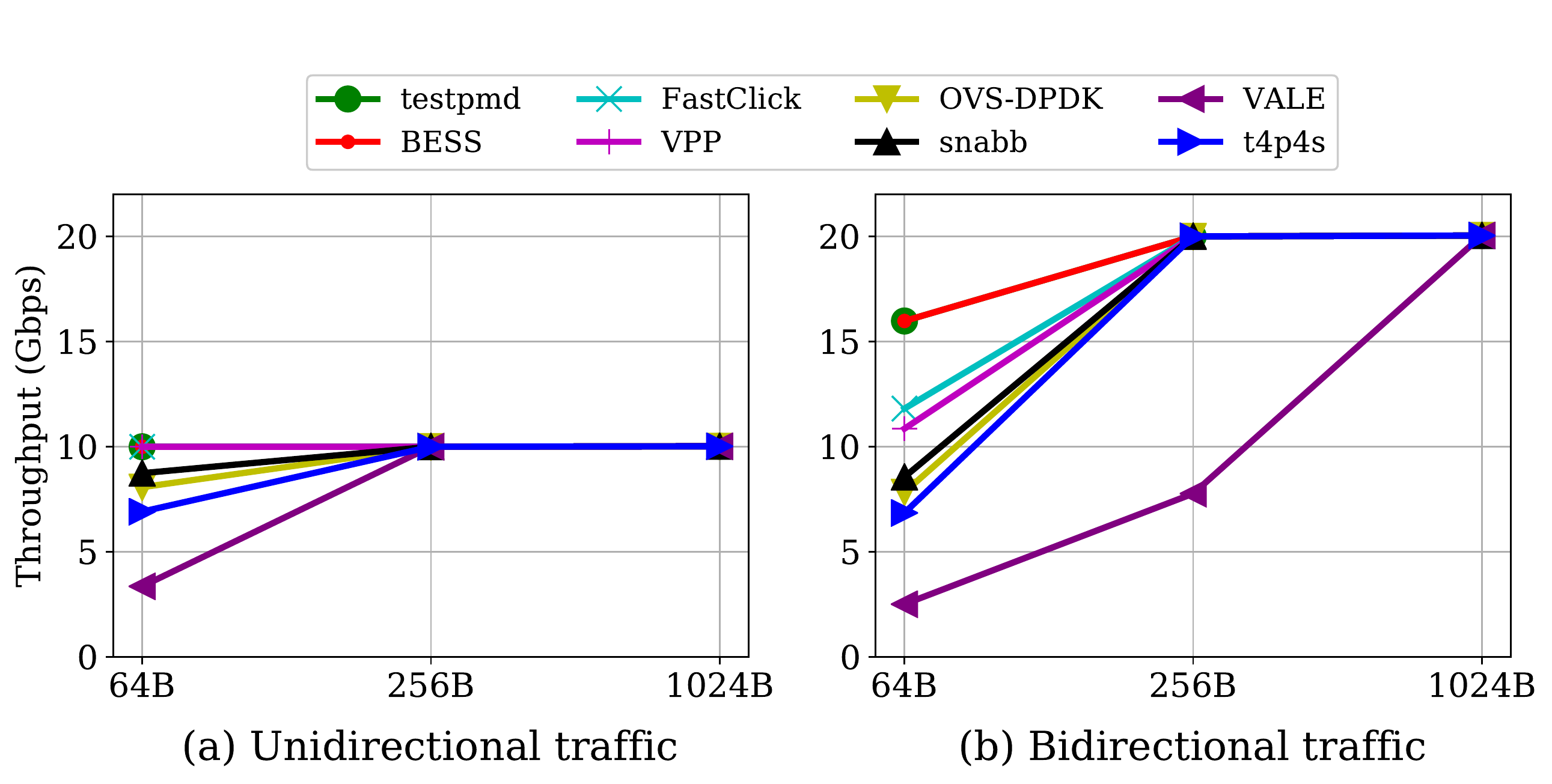}
       \caption{Throughput for the p2p scenario with unidirectional/bidirectional traffic composed of 64B, 256B, or 1024B packets.}
       \label{fig:test-p2p}
\end{figure}


\subsubsection*{p2p scenario}


Fig.~\ref{fig:test-p2p} 
shows the throughput results for the p2p scenario. 
As a simple application that only forwards packets without modification, testpmd achieves 10 Gbps with unidirectional traffic regardless of the packet size.
When considering unidirectional traffic, all the software switches saturate the 10~Gbps link with 256B and 1024B packets, proving that they are all capable of handling realistic traffic (e.g., 850B average packet size in data centers~\cite{benson2009understanding}).
For the most stressful input load with 64B packets, BESS, FastClick, and VPP still saturate the link at 10~Gbps (about 14.88~Mpps-million packets per second).
Snabb achieves only 8.74~Gbps, as staging packets in internal buffers imposes extra overhead.
OVS-DPDK achieves 8.07~Gbps due to the overhead of its match/action pipeline. As the synthetic traffic consists of identical packets, corresponding to a single flow, OVS-DPDK's flow cache does not help.
VALE switch only achieves 3.36~Gbps since, by design, it prioritizes memory isolation and therefore performs expensive packet copy operations between its ports, in addition to source MAC learning and flow table lookup.
We have tried reducing its default packet buffer size to 128B and managed to improve the forwarding rate to 14~Mpps (at the risk of server crashes).
t4p4s achieves 6.91~Gbps because it incurs the overhead of implementing multiple processing stages, including header parsing/de-parsing and flow table lookup.

Bidirectional traffic doubles the processing pressure and can therefore better reflect the forwarding capability of software switches under traffic bursts.
As shown in Fig.~\ref{fig:test-p2p}, VALE achieves 7.77 Gbps with 256B traffic, which is less than its unidirectional result (10 Gbps). This is mainly due to the twofold packet copying overhead between the physical NICs. 
All the other switches reach 20~Gbps forwarding with 256B and 1024B packets, as they do not incur heavy operations like data copying on their datapaths. 
For 64B traffic, BESS, FastClick, and VPP manage to surpass the 10~Gbps forwarding rate (which is their throughput under unidirectional traffic), but fail to reach 20 Gbps due to the overhead of their disparate internal operations. Specifically, BESS achieves the same result as the baseline testpmd application with 16~Gbps since it only performs minimal processing like collecting statistics. FastClick additionally extracts and updates packet header fields while VPP performs several verifications. The other switches achieve less throughput than unidirectional cases due to the less efficient processing pipelines and/or more complex internal operations. 
In particular, VALE suffers the most and only achieves 2.51 Gbps forwarding rate, again due to the twofold packet copying overhead.

%

\begin{figure}[!tb]
\centering
\includegraphics[width=0.48\textwidth]{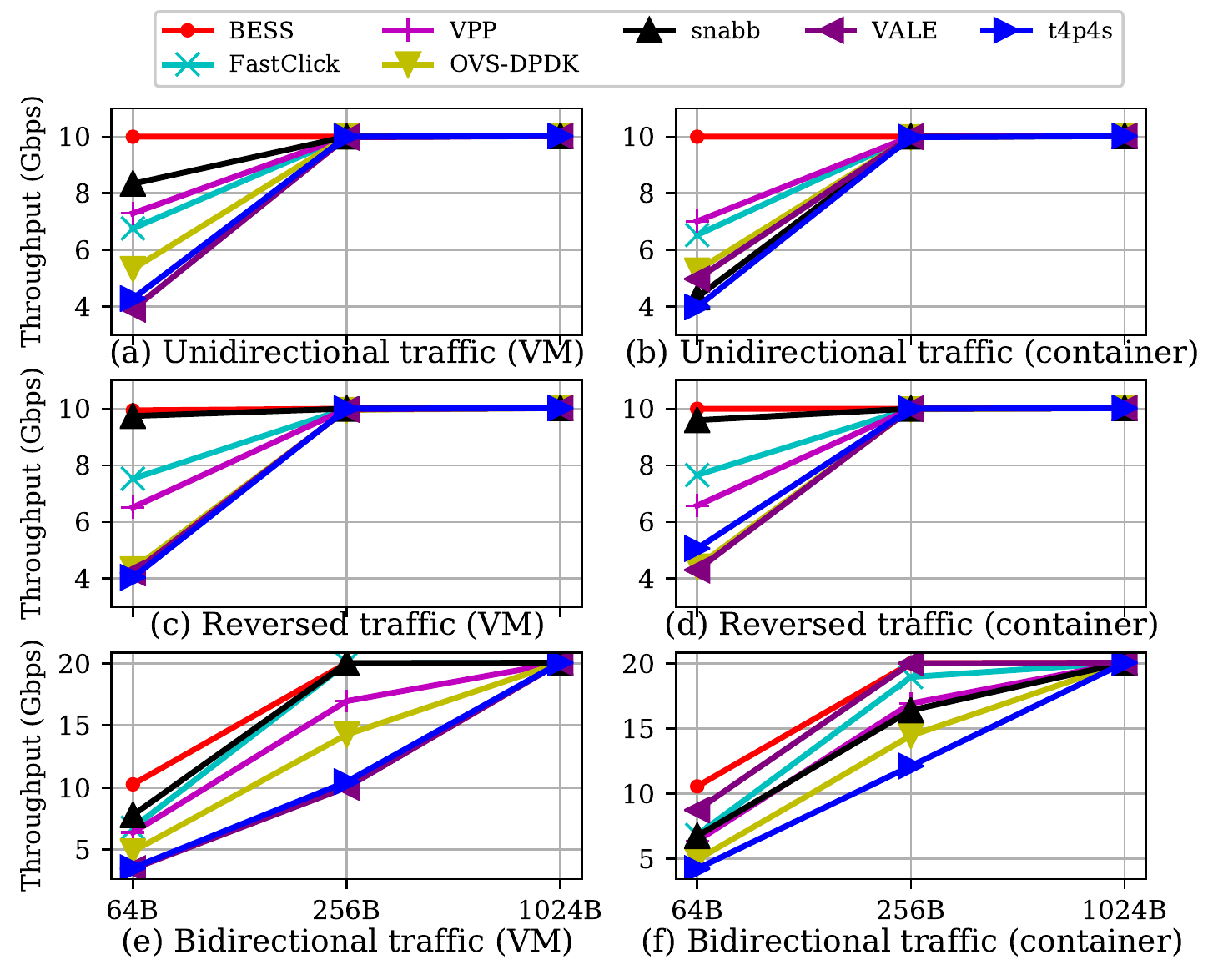}
\caption{Throughput in physical-to-virtual (p2v) scenarios with traffic generation/monitor VNF deployed in either VM or container.}
\label{fig:p2v-all}
\end{figure}

\subsubsection*{p2v scenario}\label{sec:p2v-test}

Fig.~\ref{fig:p2v-all} presents  results for the p2v scenario. 
Under unidirectional and reversed traffic, as illustrated in Fig.~\ref{fig:p2v-all}-(a)-(d), all the software switches considered in our evaluation sustain 10 Gbps under 256B and 1024B packets, showing that they are capable of forwarding more realistic traffic between physical and virtual environments.
For 64B packets, 
BESS sustains 10 Gbps in both virtual environments regardless of the overhead from the vhost-user, as the tasks it performs are very basic and simple. The performance impact of the vhost-user interface on BESS can only be understood with the more stressful bidirectional traffic discussed in the next paragraph.
VALE reaches 3.91/4.21 Gbps (VM) and 4.98/4.31 Gbps (container) with unidirectional and reverse traffic, which are better than its throughput in p2p scenario (3.36 Gbps). This is because netmap-patched ptnet driver and veth pair perform efficient packet exchange with VNFs and impose less overhead than dealing with a physical interface.
All the other switches achieve lower throughput than their p2p results. As the only difference between p2p and p2v datapaths lies in the substitution of the physical interface with a vhost-user virtual interface, therefore, these switches  experience a bottleneck in dealing with the virtualized environments as the vhost-user backend interface requires to enqueue/dequeue the virtio rings by copying packets.
For example, although FastClick and OVS-DPDK managed to sustain 10~Gbps in p2p scenario,
FastClick achieves 6.76~Gbps (VM) and 6.53~Gbps (container) with unidirectional traffic while OVS-DPDK only achieves 5.33~Gbps (VM) and 5.28~Gbps (container), respectively.
Snabb presents quite distinct results: with  reversed traffic, it achieves 9.74~Gbps (VM) and 9.60~Gbps (container); but with unidirectional traffic, it attains 8.34~Gbps for the VM test and merely 4.33~Gbps for the container test (approximately 48\% difference).
We believe the low throughput with unidirectional traffic when containers are used comes from the compatibility issue between Snabb switch and virtio-user frontend. The same issue also occurs in the loopback test discussed below.
Snabb also achieves much higher throughput with reversed traffic, the root cause for this is still under investigation.

Fig.~\ref{fig:p2v-all}-(e) and (f) illustrate the measured throughput with bidirectional traffic. For 256B and 1024B packets, BESS still sustains line rate, i.e., 20~Gbps, but the impact of vhost-user is noticeable for the other switches. Indeed, VPP, OVS-DPDK, and t4p4s fail to saturate 20~Gbps in both virtual environments while FastClick and Snabb fail in the container test, in contrast to the 20 Gbps they achieved in the bidirectional p2p test.
In particular, although Snabb outperforms VPP in the VM test (20~Gbps vs.\ 16.96~Gbps), it only achieves 16.38~Gbps in the container test and is outperformed by VPP (16.91~Gbps). By correlating with the p2v unidirectional container test, we believe the abnormally low throughput causes this degradation.
VALE only gets 10~Gbps rate in the VM test due to the extra overhead imposed by the VALE instance. The real throughput is expected to be much higher if ptnet interface is compatible with netmap's pkt-gen. So the results here only represent a lower bound, which can be verified in the container case (20~Gbps). 
For 64B traffic, BESS achieves 10.26~Gbps for VM test and 10.54~Gbps for container test respectively, which are much lower than its bidirectional p2p test result (16~Gbps), further illustrating the impact of vhost-user.
VALE attains 8.72~Gbps and outperforms the other 5 switches in the container test, thanks to the efficiency of the patched veth pair. Although it only achieves 3.47~Gbps in the VM test because of the overhead of one extra in-path VALE instance, this result is still higher than its bidirectional p2p throughput (2.51~Gbps), which indirectly reflects the efficiency of netmap's ptnet mechanism.
The other switches follow the same trend as the unidirectional tests. Note that, except for the special cases of VALE and Snabb, the other switches achieve similar throughput in both VM and container tests.

\subsubsection*{v2v scenario}
\begin{figure}[!tb]
\begin{center}
\includegraphics[width=0.47\textwidth]{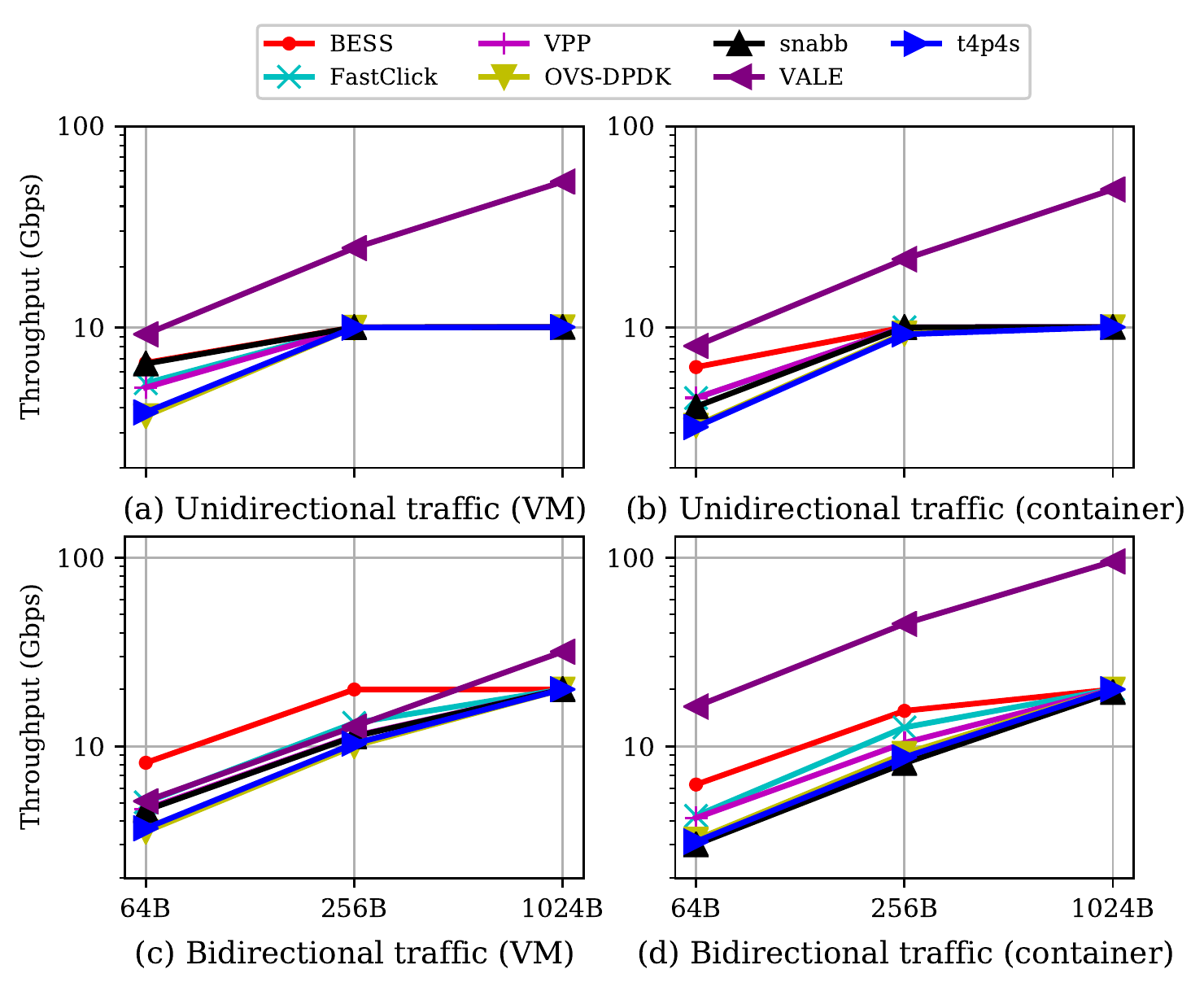}
\caption{Throughput in virtual-to-virtual (v2v) scenarios with different traffic patterns and virtual environments. Note that the y-axis shows the measured throughput in log scale.}
\label{fig:v2v-all}
\end{center}
\end{figure}

Results of the v2v throughput test with both VMs and containers are reported in Fig.~\ref{fig:v2v-all}.
In the unidirectional test, as illustrated in Fig.~\ref{fig:v2v-all}-(a) and (b), most of the switches achieve at least 10~Gbps throughput with 256B and 1024B traffic. 
As no physical NICs are involved in the v2v scenario, the achievable throughput is only limited by the system bus of our server.
Indeed, VALE achieves 24.80~Gbps (VM) and 21.82~Gbps (container) with 256B traffic; for 1024B traffic, its throughput even reaches 52.95~Gbps (VM) and 48.53~Gbps (container).
The throughput of BESS, FastClick, VPP, and Snabb are upper-bounded by 10~Gbps due to the rate limitation of pktgen-DPDK on vhost-user interfaces. The actual throughput should be higher than this.
In our previous work~\cite{zhang2019comparing}, we used MoonGen inside a VM in the v2v scenario to accomplish flexible traffic generation beyond 10~Gbps. However, MoonGen does not work with virtio-user frontend interfaces inside the container namespace. Therefore, we use pktgen-DPDK in this scenario to provide a fair comparison for both virtual environments. A full-fledged traffic generator combining the advantages of MoonGen and pktgen-DPDK is left for future work.
In fact, the current limitation of pktgen-DPDK does not affect our observation on 64B unidirectional traffic as no switch exceeds 10~Gbps.
In this case, VALE is still the most performant switch and achieves 9.24/8.08~Gbps for VM/container. Compared with its corresponding p2v result (3.91/4.98~Gbps), it is clear that VALE is more efficient in both VM and container networking. This is mostly because of the efficiency of the ptnet/veth zero-copy packet I/O mechanism. 
All the other switches achieve throughput lower than 6.6~Gbps and experience throughput degradation compared to p2v tests due to the extra overhead introduced by one additional vhost-user interface.
In particular, BESS achieves 9.67~Gbps (VM) and 9.34~Gbps (container) respectively, which are less than the 10~Gbps throughput it achieved in p2p and p2v unidirectional tests. 
Snabb still achieves distinct throughput with different virtualization environments in the v2v unidirectional test, with 6.59~Gbps (VM) and 4.03~Gbps (container), due to its incompatibility issue with virtio-user frontend inside Docker.

Figs.~\ref{fig:v2v-all}-(c) and (d) illustrate the measured bidirectional throughput for VNFs running in VMs and containers. 
VALE presents the highest throughput in the container test with 16.24~Gbps for 64B traffic, 44.65~Gbps for 256B traffic, and 95.55~Gbps for 1024B traffic. Even in the VM test, it still outperforms most of the other switches with 5.11~Gbps (64B), 12.72~Gbps (256B), and 31.74~Gbps (1024B), despite the limitation of ptnet and netmap's pkt-gen (as explained in prior sections).
These results again demonstrate the efficiency of netmap's ptnet/veth in dealing with VNF networking.
Other switches achieve at least 20~Gbps throughput with 1024B traffic due to the limitation of pktgen-DPDK. 
For 64B and 256B traffic, BESS achieves the highest throughput in the VM test with 8.17~Gbps and 20~Gbps and outperforms the others. It also presents the second-best results after VALE, thanks to its simple internal operations.
FastClick, OVS-DPDK, VPP, Snabb, and t4p4s also exhibit sightly lower throughput with bidirectional traffic compared to their unidirectional results, mainly because of the doubled internal processing and packet copying operations through vhost-user interfaces.
Note that Snabb has the worst throughput in container tests due to its compatibility issue with containers as explained before. In the VM test, it can still outperform OVS-DPDK and t4p4s as they suffer from the greater overhead of their Match/Action pipelines.

\subsubsection*{Loopback scenario}

\begin{figure}[!tb]
\begin{center}
\includegraphics[width=0.5\textwidth]{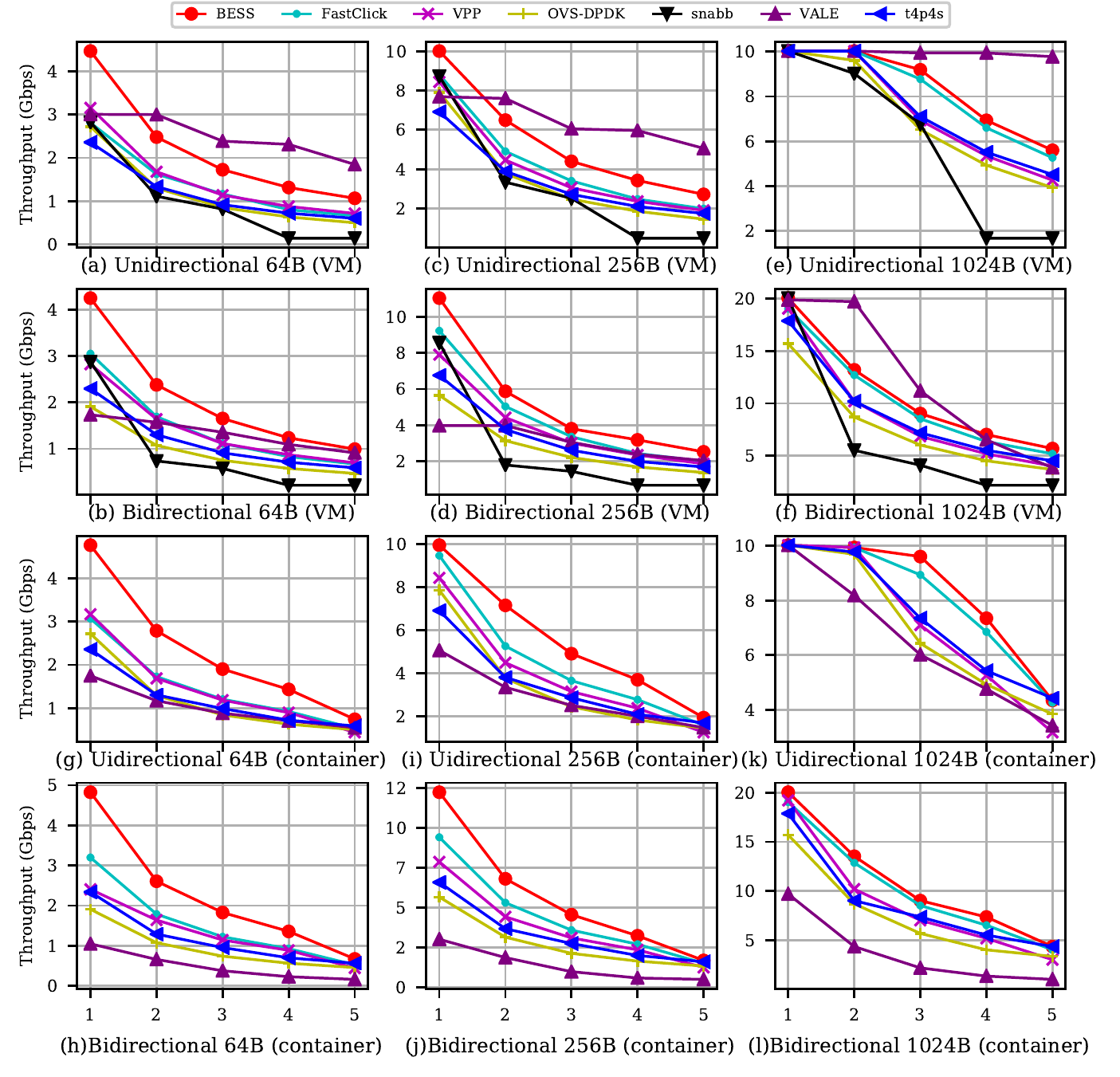}
\caption{Throughput measurement of the loopback scenario with VNFs deployed in either VMs or containers. The x-axis represents the service chain length (1-5) and the y-axis represents the measured throughput (in Gbps).}
\label{fig:lo}
\vspace{-0.8cm}
\end{center}
\end{figure}

Figs.~\ref{fig:lo}-(a)-(f) illustrate the throughput for the loopback scenario with different input traffic and VNFs deployed inside QEMU/KVM VMs.
In particular, results for unidirectional traffic are shown in Figs.~\ref{fig:lo}-(a), (c), and (e). Intuitively, the measured throughput decreases as the length of the service chain grows since packets are steered back and forth through the software switch between the successive VNFs. Since we run each software switch with a single core,  throughput degradation becomes more obvious as more VNFs are appended to the service chain. Besides, we configure VNFs to update destination MAC addresses. Although this overhead is tiny, it also accumulates with longer chains.
As can be seen, BESS yields the highest throughput with a single VM. However, it is outperformed by VALE with two or more chained VNFs. This is mainly because BESS needs to perform an increasing number of packet copies as the number of VMs increases. Even though VALE still needs to copy packets between the physical ports, this overhead is compensated by the efficient VM network I/O of ptnet.
As shown in Fig.~\ref{fig:lo}-(e), VALE manages to sustain 10~Gbps for unidirectional traffic with 1024B packets even with four chained VMs. BESS still outperforms other switches because of its simple processing.
Other switches achieve lower throughput due to the overhead (mainly packet copies) imposed by vhost-user. Snabb still outperforms OVS-DPDK and t4p4s with a single VNF, but becomes overloaded with two or more VNFs. When the service chain length reaches 4, its packet rate becomes constant, as the workload is too much to be handled with a single core. This is expected to be alleviated by allocating multiple cores to share the workload.
The throughput of FastClick and VPP are very close with 64B and 256B traffic, as shown in Fig.~\ref{fig:lo}-(a) and (c). But with 1024B traffic, FastClick achieves almost the same throughput as BESS, and outperforms VPP by a large margin. This is because FastClick has simpler internal processing than VPP, which becomes more obvious with less stressful traffic. With more intensive traffic in the case of 64B packets, the optimized vector processing pipeline of VPP makes it more advantageous than FastClick~\cite{vppcommag2018,vppcomnet2019,barach2018batched}.
For bidirectional traffic, as illustrated in Figs.~\ref{fig:lo}-(b), (d) and (f), all the tested software switches present decreasing throughput as the service chain length grows, as for the unidirectional traffic case.
In particular, VALE experiences significant performance degradation, especially for 64B and 256B traffic. For 1024B traffic, its performance begins to drop when the service chain length is greater than~2. This is mainly due to the doubled packet copying overhead between the neighboring ports of sequential VALE instances. 
The throughput of Snabb drops by 4~Gbps with 1024B traffic when two VNFs are added to the chain, since bidirectional traffic already imposes insurmountable overhead.  
Other software switches are only able to sustain throughput comparable to their counterparts in unidirectional experiments, due to the doubled number of packet copying operations between their vhost-user backend and the virtio-pci interfaces of VMs, as well as the doubled processing overhead of the chained VNFs inside VMs.

Figs.~\ref{fig:lo}-(g)-(l) present the results for the loopback test scenario with VNFs deployed inside Docker containers. As explained in Sec.~\ref{sec:los}, experiments for Snabb are omitted here since we cannot configure its datapath with Docker containers. Its results for VM tests can give an approximate estimation for container scenarios.
In general, all the switches follow the same trend as the VM case and their throughput decreases as the chain length grows due to the accumulating processing and forwarding overhead.
In particular, VALE fails to achieve the same efficiency as in the VM case and is outperformed by other switches in most cases. This result demonstrates that veth, though it works in a zero-copy manner, is not as efficient as ptnet on service chains. As VALE achieved similar throughput with veth and ptnet interfaces in the p2v and v2v scenarios, we speculate that the bottleneck comes from the VALE instance inside each container since interconnecting two veth interfaces imposes non-negligible switching overhead. 
All the other switches achieve comparable throughput to that in their corresponding VM scenario. Thus, we conclude that most of the software switches considered in our study can achieve similar throughput networking VNFs inside either QEMU-based VMs or Docker containers.

\subsection{Latency Tests}\label{sec:ltest}

\subsubsection*{p2p scenario}

\begin{figure}[!tb]
  \centering
       \includegraphics[width=0.48\textwidth]{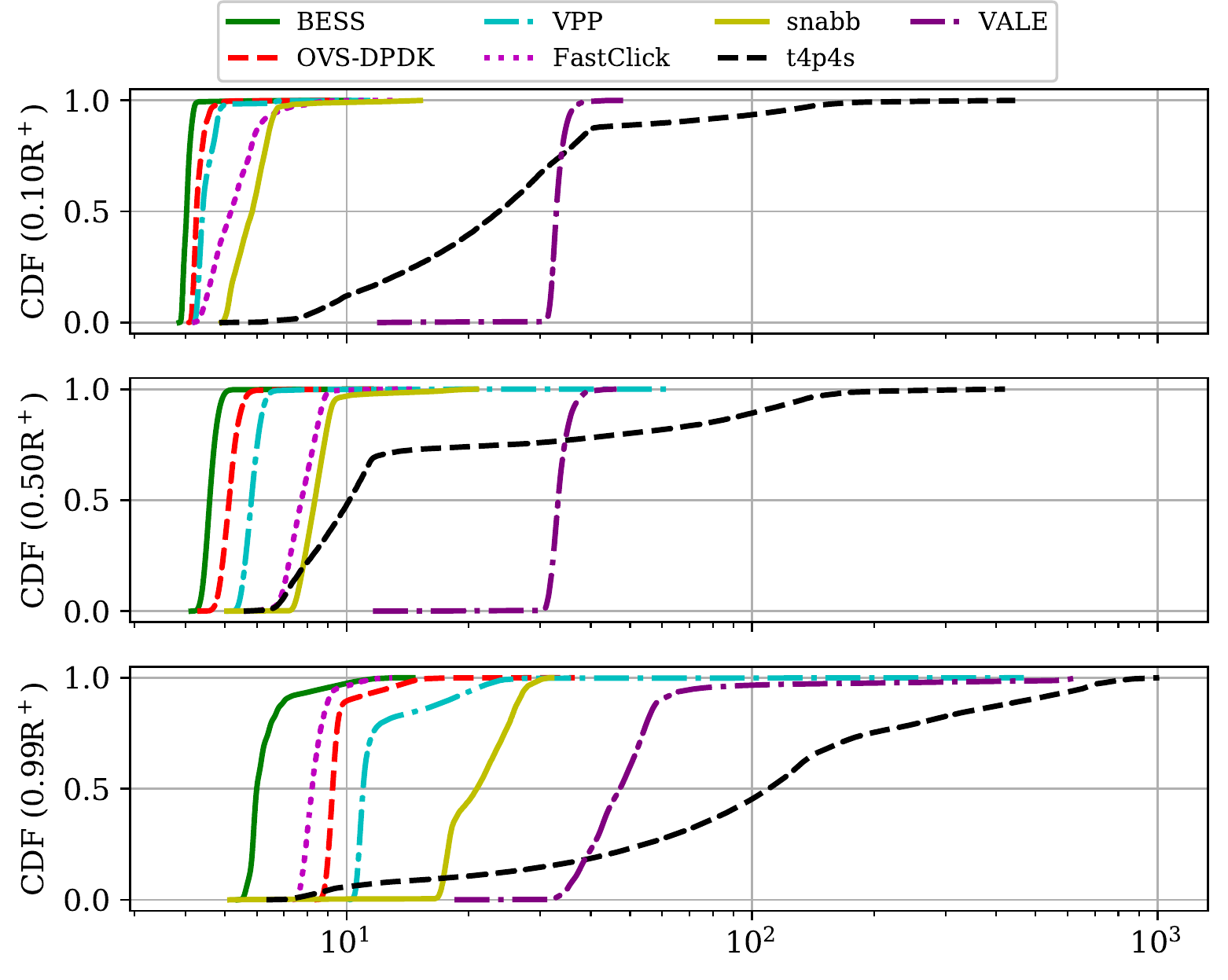}
       \caption{The CDFs of RTT latency (in $\mu s$) for all the software switches in the p2p test scenario. These results are obtained using MoonGen's hardware time-stamping. As the RTT ranges from 4 $\mu$s up to 1000 $\mu$s, we plot x-axis in log scale for better visualization.}
       \label{fig:test-p2p-latency}
\end{figure}

Fig.~\ref{fig:test-p2p-latency} shows the CDF of the measured RTT latency with different input rates. As explained in Sec.~\ref{subsec:exp:latency1}, we define $R^+$ to be the forwarding rate we obtained in throughput tests, it is thus different for each software switch.
In general, most of the switches experience smaller RTT as the input load reduces from $0.99R^+$ to $0.10R^+$, since their datapaths become less congested.
Under the 0.99$R^+$ rate, the RTTs for t4p4s are very high and unstable, showing its instability under high loads. Since the other DPDK-based switches encounter no such problems, we believe this is due to the inefficiency of the t4p4s internal pipeline. The hardware abstraction layer of t4p4s presents a trade-off between performance and platform independence, and the level of abstraction could be re-factored to enhance performance. 
VALE also presents the second-worst RTT at 0.99$R^+$ because it combines polling and interrupts for packet I/O, which is less stable compared to pure busy-waiting for packet I/O. 
Snabb latency is also quite high mainly because the internal buffers of its in-path modules prolong the packet journey under high input load.
BESS, FastClick, OVS-DPDK, and VPP exploit polling and batching to achieve low latency.
Under 0.50$R^+$ load, the RTTs of t4p4s are almost 10 times lower as the input traffic is almost 50\% less stressful. 
VALE presents the worst latency with most of its RTTs distributed between 30-40~$\mu$s. This is because, under a low input rate, VALE automatically switches to interrupt processing to save CPU cycles for other tasks. The interrupt-based packet I/O is usually more time-consuming than poll-mode due to interrupt propagation latency, context switch latency, and cold data/instruction caches. 
BESS, Snabb, VPP, and OVS-DPDK exhibit better RTT reduction, as the decreased input rates make their processing pipelines less congested.
Under 0.10$R^+$ load, 
t4p4s achieves a worse RTT distribution than in the 0.50$R^+$ test. This is a consequence of the extra delay in constituting batches under low input rates. 
VALE presents almost the same RTT distribution as under $0.5R^+$ load, since it processes packets with interrupts. 
The other switches present better latency with their RTTs mostly distributed between 4 and 7~$\mu$s, thanks to the low input rate and efficient poll-mode processing.

\subsubsection*{Loopback scenario}

\begin{figure*}[!tb]
	\centering
	\begin{subfigure}[t]{0.32\textwidth}
		\centering
		\includegraphics[width=1.05\textwidth]{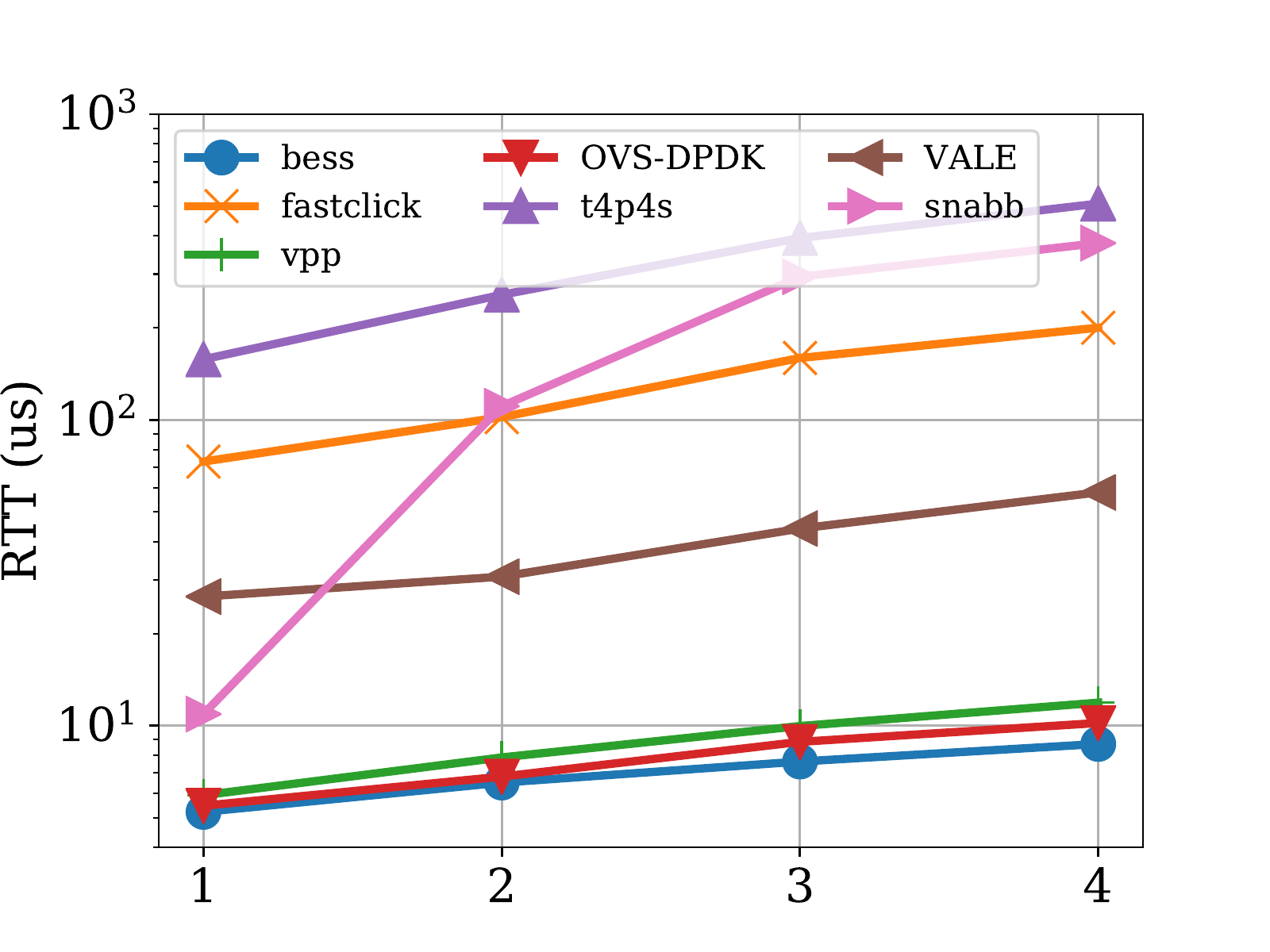}
		\caption{0.10$R^+$ with VNFs deployed in VMs}
	\end{subfigure}
	~
	\begin{subfigure}[t]{0.32\textwidth}
		\centering
		\includegraphics[width=1.05\textwidth]{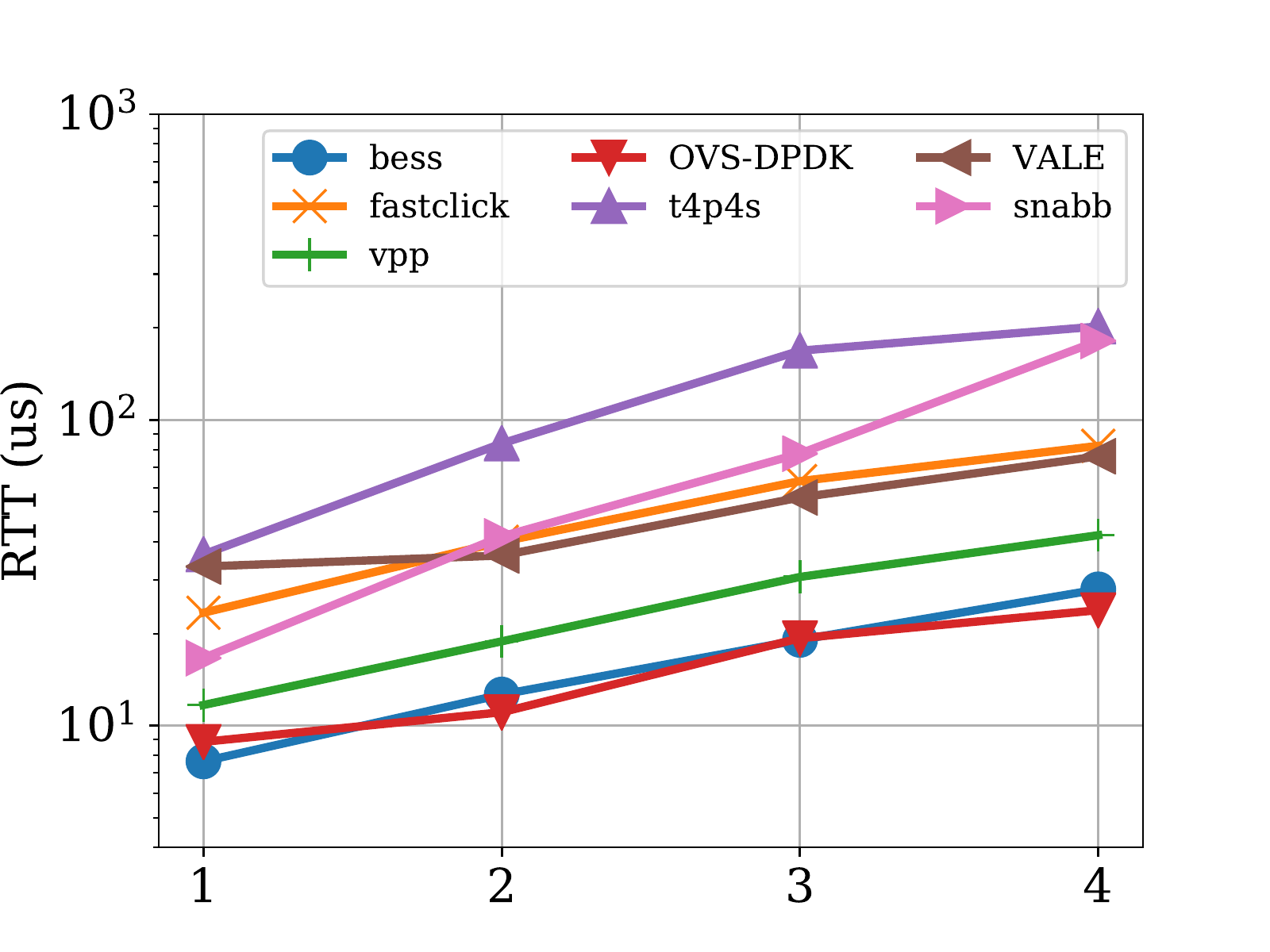}
		\caption{0.50$R^+$ with VNFs deployed in VMs}
	\end{subfigure}
	~
	\begin{subfigure}[t]{0.32\textwidth}
		\centering
		\includegraphics[width=1.05\textwidth]{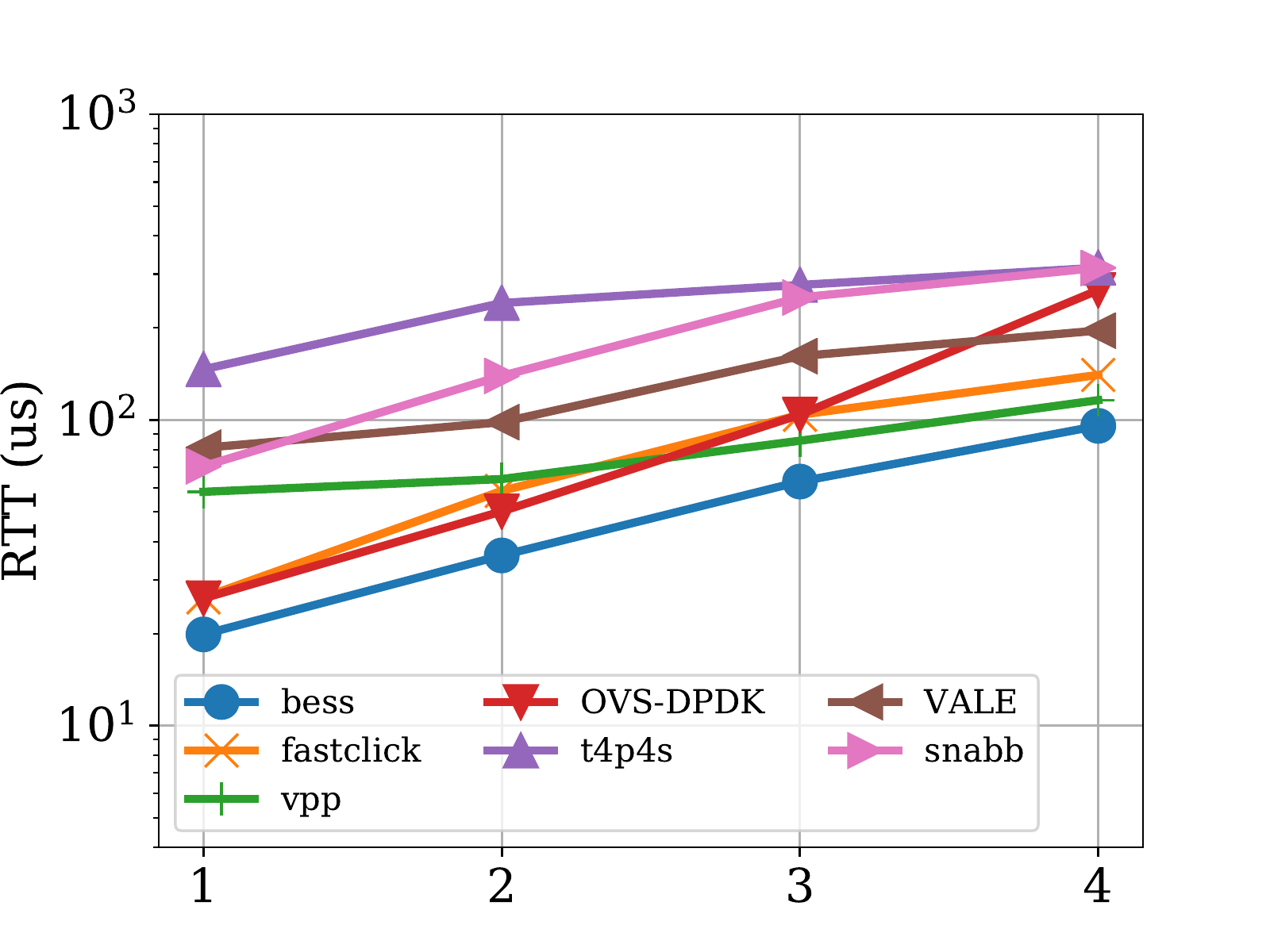}
		\caption{0.99$R^+$ with VNFs deployed in VMs}
	\end{subfigure}

	\begin{subfigure}[t]{0.32\textwidth}
		\centering
		\includegraphics[width=1.05\textwidth]{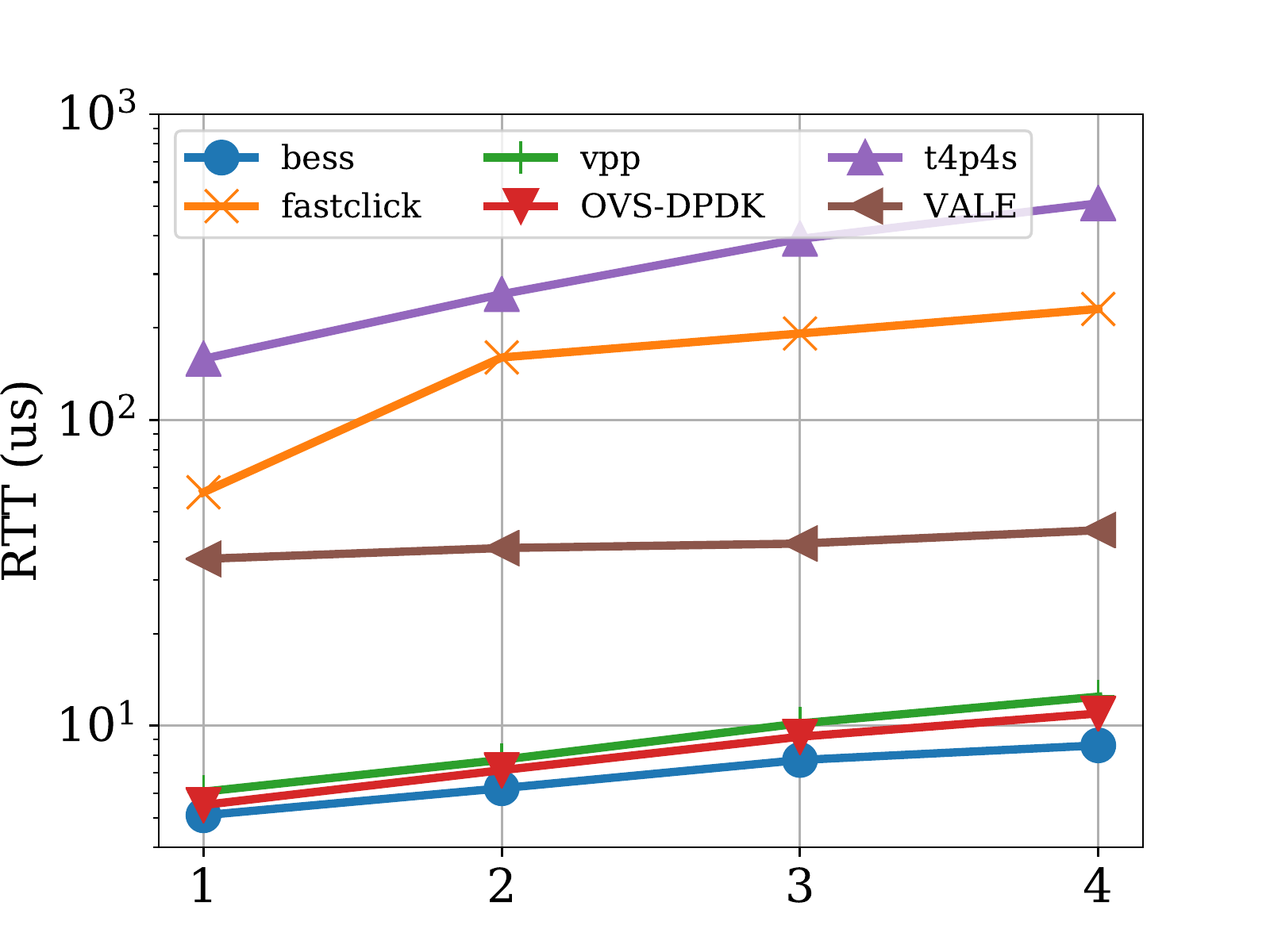}
		\caption{0.10$R^+$ with containerized VNFs}
	\end{subfigure}
	~
	\begin{subfigure}[t]{0.32\textwidth}
		\centering
		\includegraphics[width=1.05\textwidth]{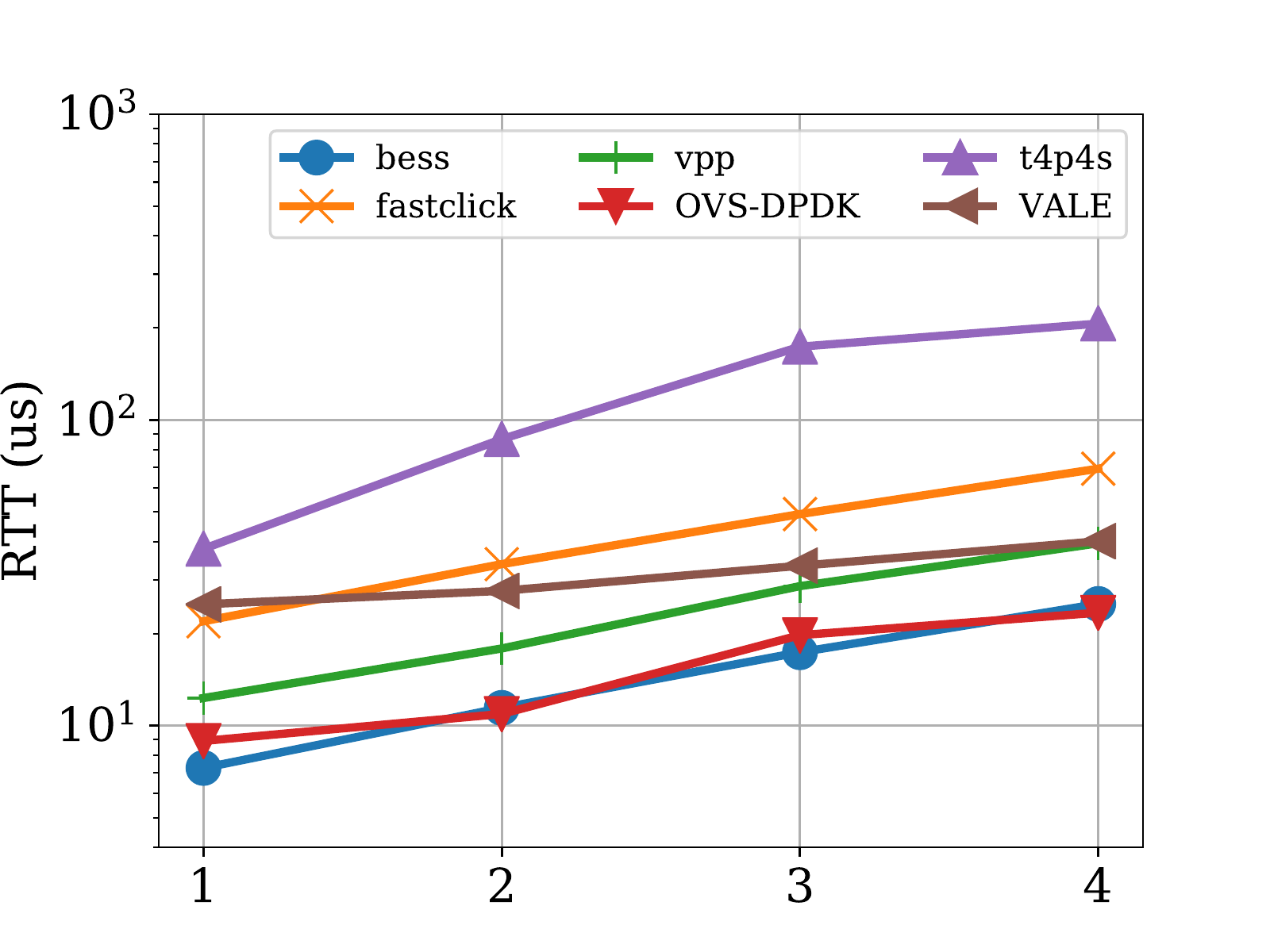}
		\caption{0.50$R^+$ with containerized VNFs}
	\end{subfigure}
	~
	\begin{subfigure}[t]{0.32\textwidth}
		\centering
		\includegraphics[width=1.05\textwidth]{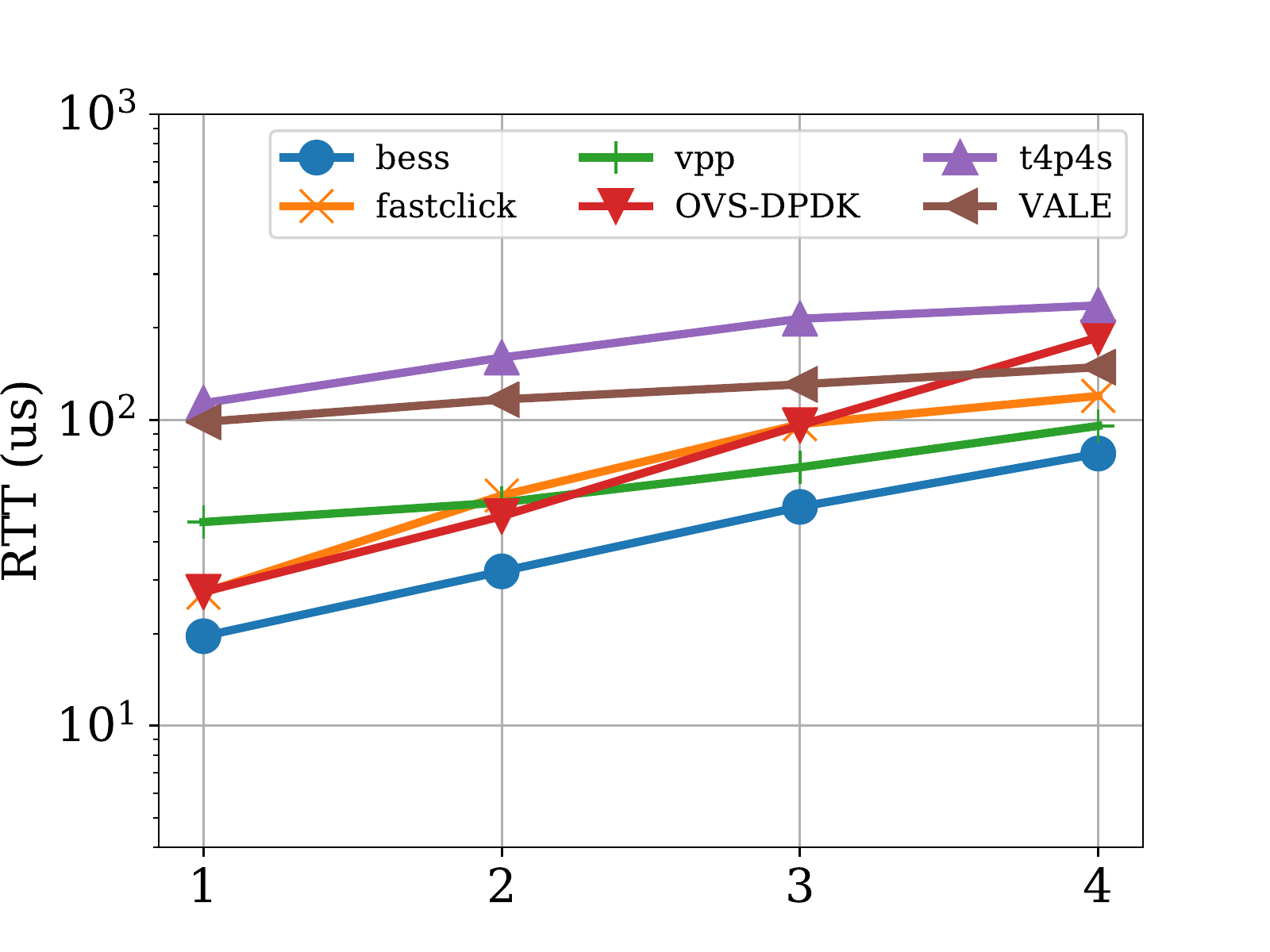}
		\caption{0.99$R^+$ with containerized VNFs}
	\end{subfigure}
	\caption{Latency measurement (as RTT in $\mu$s) of the loopback scenario with VNFs in either VMs or containers}
	\label{fig:lo-rtt}
\end{figure*}

Fig.~\ref{fig:lo-rtt} shows the latency test results in the loopback scenario with service chain length varied from 1 to 4. 
For the sake of space, instead of plotting the CDFs, we only illustrate the average RTTs for each software switch with VNFs deployed in VMs, as shown in Figs.~\ref{fig:lo-rtt}-(a)-(c), and in containers, as illustrated in Figs.~\ref{fig:lo-rtt}-(d)-(f).
We omit the results for Snabb in the container test due to its compatibility issue with virtio-user as explained previously.
In general, for all the switches we test, the latency with 0.99$R^+$ load is always higher than with 0.50$R^+$ load. This is as expected, since $R^+$ is only the average throughput and the actual forwarding rate of each software switch fluctuates around this. Consequently, an unstable software switch might fail to sustain 0.99$R^+$ in a specific period, causing data path congestion and packet loss. Such a situation rarely happens under 0.50$R^+$ load. 
Another significant result is the impact of batch processing of some software switches since, at a low input rate, time has to be spent to wait for new packets to complete a batch, thus increasing overall latency.
As shown in Figs.~\ref{fig:lo-rtt}-(a) and (d), latency under 0.10$R^+$ load is higher than under 0.50$R^+$ for t4p4s, Snabb, and FastClick, mainly due to their internal strict batch processing. Although FastClick flushes its packet buffer by default, its internal batch processing mechanism still imposes higher processing latency at low input rates. 
We did not observe the same effect for FastClick and Snabb in the p2p test because the batch effect could not accumulate as the loopback scenario with packets traversing FastClick multiple times.
All the other switches do not encounter this issue since they dynamically adjust the batch size, and their RTTs do not increase so much as the service length grows longer.
In all cases, t4p4s presents the worst latency, reflecting the inefficiency of its processing pipeline. BESS achieves optimal latency in all cases because it has the simplest pipeline and only performs minimal processing on each packet.
Note that Snabb presents a huge latency leap from the 2-VNF VM test since it becomes overloaded at this point and fails to keep up with input traffic. The same phenomenon was observed in its throughput test. OVS-DPDK also experiences a relatively large leap in the 4-VNF case, for both VMs and containers, due to similar reasons.
In general, most of the switches achieve similar RTTs with service chains of varied length deployed in both VMs and containers.

\subsubsection*{v2v scenario}

\begin{figure}[!tb]
  \centering
       \includegraphics[width=0.5\textwidth]{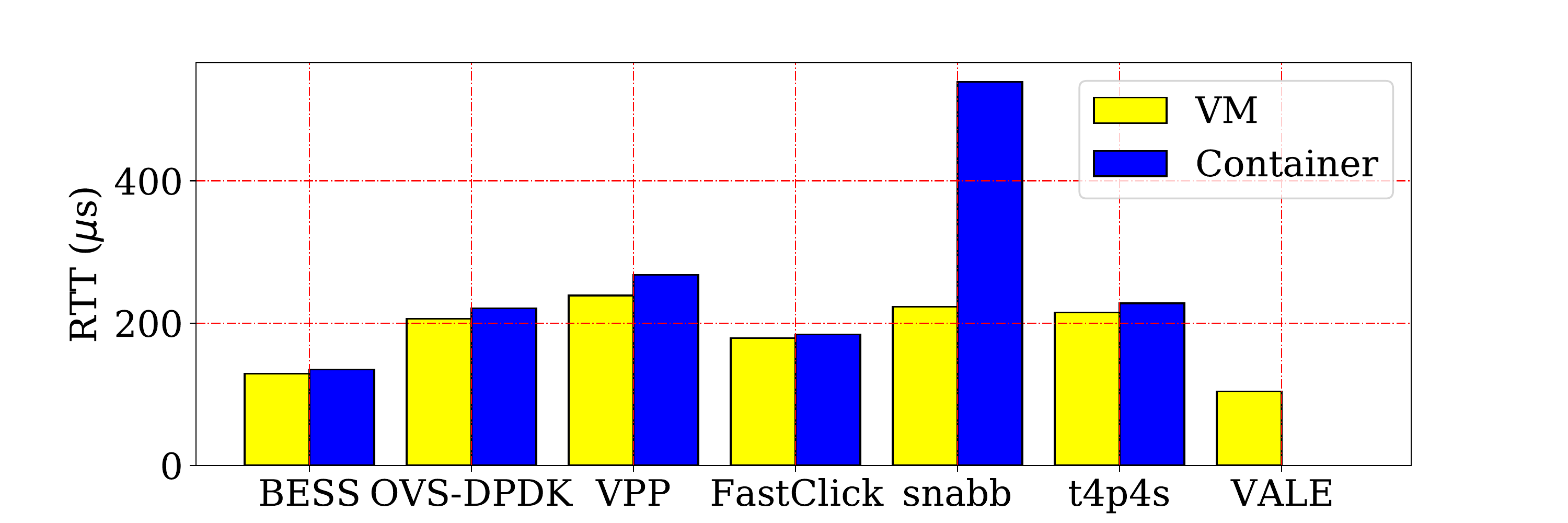}
       \caption{RTT latency (in $\mu s$) for the v2v scenario based on pktgen-DPDK latency test function.}
       \label{fig:test-v2v-latency}
\end{figure}

\begin{table*}[!tb]
\scriptsize
  \centering
  \caption{Summary of use cases for each software switch}
  \vspace*{-0.1cm}
  \label{tab:usecases}
  \begin{tabular}{|l|l|l|}
	\hline 	 & {\bf Best use cases} &  {\bf Remarks} \\
  	\hline\hline
	BESS & Forwarding between physical NICs and containers & Chaining of containerized VNFs  \\
 	\hline
	Snabb & Fast deployment, runtime optimization & Bottlenecked with multiple VNFs \\
	\hline
	OVS-DPDK & Stateless SDN  deployments & Supports OpenFlow protocol \\
	\hline
	FastClick &  VNF chaining & Flexible live migration, high latency at low workload \\
	\hline
	VPP & VNF chaining  & Flexible live migration  \\
	\hline
	VALE & VNF chaining with high workload &  Limited traffic classification and live migration capability \\
 	\hline
	t4p4s & Stateful SDN deployments & Supports P4 semantics \\
	\hline
\end{tabular}
\end{table*}

As MoonGen's rate control and time-stamping feature are tightly coupled with physical NICs, we cannot perform the same latency test to measure RTTs with different input rates.
To provide a fair comparison of the processing latency for the software switches, we opt for pktgen-DPDK and explore its  latency measurement feature that works inside both VMs and containers. The results of the v2v latency tests are shown in Fig.~\ref{fig:test-v2v-latency}. 
VALE outperforms other switches in terms of latency (only 100~$\mu$s), as was previously observed for the v2v throughput test. BESS, FastClick, VPP, t4p4s, and OVS-DPDK achieve very similar latencies for both VMs and containers, as they all use vhost-user to interconnect VNFs. BESS achieves the best RTT among them due to its simplicity, which is coherent with throughput tests. Snabb presents quite a high RTT with containerized VNFs. We believe this is due to the compatibility issue between Snabb's vhost-user backend and the container-side virtio-user frontend, as explained in Sec.~\ref{sec:p2v-test}.
While ptnet requires two packet copying operations between VALE ports, solutions based on vhost-user have to incur four copies on virtio rings.

\subsection{Best Match Between Use-Cases and Switches}
Based on the previous experimental results, we can make the following remarks on possible use cases for the considered switches. These remarks complement the taxonomy previously presented in Table~\ref{tab:taxonomy}.

BESS achieves both high throughput and low latency in p2p, p2v, and 1-VNF loopback scenarios for VMs.
It also achieves optimal performance for all the container-based loopback tests.
It is a viable choice to switch traffic between physical NICs and one or multiple paralleled VMs, as well as to steer packets for containerized service chains.

Snabb performs well in most cases but suffers from overload in the loopback scenario with a chain of more than 3 VNFs. We also failed to find a solution to implement the loopback test with containers due to a compatibility issue.
It is easier to deploy than other solutions based on DPDK or netmap and is thus a good choice when the time-to-production of specific applications is critical.

OVS-DPDK and t4p4s have the advantage of supporting OpenFlow and P4, respectively, and are thus the only solutions that work with third-party SDN controllers and newly introduced protocols. OVS-DPDK appears the best option for a stateless SDN scenario, while t4p4s is preferable when some state is required (e.g., for a firewall).


FastClick and VPP have good performance in all scenarios and simplify service migration thanks to the isolation bestowed by virtio-pci/virtio-user frontend and vhost-user backend. Moreover, unlike BESS, they are compatible with newer hypervisor versions and can, therefore, be used to build both linear and parallel NFV environments with reasonable trade-offs. Compared to FastClick, VPP might be preferred when the latency is the primary concern since it generally has lower RTT and avoids severe latency degradation at low input rates (e.g., $0.1R^+$).

Finally, VALE, augmented by ptnet passthrough and cross-namespace veth, achieves relatively high throughput in v2v and the loopback scenarios. It is, therefore, well-suited to construct linear service chains in environments with high workloads. On the other hand, as ptnet and veth are highly dependent on the host netmap module, they do not have the same level of memory isolation as the virtio para-virtualized driver. Thus, migrating its VNFs may require synchronization at the host level. Another caveat is that VALE, as a simple Ethernet switch, has limited capability for traffic classification compared to the other solutions and may require enhancement to support more advanced traffic processing and forwarding features.


\section{Conclusion}\label{conclusion}
The emergence of high-speed packet I/O frameworks and the proliferation of NFV have given rise to intense research on the design of software switches running on COTS servers. Many different designs have been proposed and implemented to route traffic between NICs and VNFs on NFV platforms. In this paper, we have sought to improve understanding of the throughput and latency performance of these alternative designs by defining a performance measurement methodology and providing sample results for 7 state-of-the-art proposals.

The methodology is based on four test scenarios; physical to physical (p2p); physical to virtual (p2v); virtual to virtual (v2v); loopback (with multiple sequentially chained VNFs); designed to explore the performance of typical NFV configurations, where traffic is forwarded between multiple physical and virtual interfaces. In the interest of reproducibility, the paper describes the experimental setup in detail, including specifications of tested software and hardware versions and the packet generation and monitoring tools used.
In our evaluation, all the VNFs are hosted in two most popular virtualized environments, namely virtual machines (QEMU/KVM) and containers (Docker).
The measurement results reveal that no single switch prevails in all scenarios. Such a result was expected, given the different design objectives of the considered software switches, but is a useful reminder that the best switch choice depends significantly on the intended NFV context. The presented results and related discussion enable a more informed choice and should guide the design of potential enhancements to relieve identified bottlenecks.

It is worth noting that the present wide-range comparison has required considerable effort to understand the details of the considered switches and set up and conduct the experiments. Therefore, we hope that other researchers will be able to profit from our experience, by further exploring the performance dimensions of existing and emerging software switches and refining the evaluation methodology.

\section*{Acknowledgments}
This work has been carried out at LINCS (\url{http://www.lincs.fr}) and benefited from the support of NewNet@Paris, Cisco's Chair ``{\sc Networks for the Future}'' at Telecom ParisTech (\url{https://newnet.telecom-paristech.fr}). The authors are very grateful for the collaboration with Massimo Gallo, who coauthored a preliminary version of this work.






\bibliographystyle{elsarticle-num-names}
\bibliography{stack}
%
%
%

%


%

\appendix

\section{Switch configurations}
\label{sec:config}

To enhance reproducibility, we provide the most critical configuration details for each test scenario.
\subsection*{p2p scenario}
Each switch requires a unique configuration to implement the p2p scenario. We only show the most critical configuration snippet for each design.
For BESS, we composed a configuration script in which physical interfaces are hooked to the \emph{bessd} daemon process with the built-in {\tt PMDPort} module. Physical queues (input/output) of the hooked interfaces are instantiated using {\tt QueueInc/QueueOut} modules. Packet forwarding is implemented by linking different queues with the right arrow:
\begin{lstlisting}
inport::PMDPort(port_id=0,...)
outport::PMDPort(port_id=1,...)

in0::QueueInc(port=inport, qid=0)
out0::QueueOut(port=outport, qid=0)

in0 -> out0
\end{lstlisting}

For FastClick, we compose a similar configuration file with {\tt FromDPDKDevice}/{\tt ToDPDKDevice} modules that hook and link two physical interfaces as follows: 
\begin{lstlisting}
FromDPDKDevice(0,...)->ToDPDKDevice(1,...)
\end{lstlisting}
Note that it is easier just to whitelist the physical interfaces using DPDK ``-w" option. 

For t4p4s, we select its l2fwd application which learns the source MAC address and forwards packets according to a predefined flow table. The table is configured with ``destination MAC address/output port" as Match/Action fields. 

For VPP, we specify the PCI addresses of the interfaces in the configuration file. We interconnect the ports with the \emph{l2patch} function, as this is functionally equivalent to the configuration of other switches:
\begin{lstlisting}
test l2patch rx port0 tx port1
test l2patch rx port1 tx port0
\end{lstlisting}

For Snabb switch we similarly write a custom module that hooks the ports by PCI addresses and recompile the Snabb software to make the module executable. Inside the module, we start a new configuration object and instantiate two logical port 
``apps" using the PCI port addresses which are then interconnected through the ``link" method:

\begin{lstlisting}
local c = config.new()
config.app(c, "nic1",...,{pciaddr = pci1})
config.app(c, "nic2",...,{pciaddr = pci2})
config.link(c, "nic1.tx -> nic2.rx")
\end{lstlisting}

For OVS-DPDK, we configure a new bridge and attach the physical interfaces to it by specifying their PCI addresses using the {\tt OVS-vsctl} command. Then we populate the flow table with direct forwarding rules between the interfaces using the {\tt OVS-ofctl} command.

For VALE, we need to unload the {\tt ixgbe} kernel module and load its netmap counterpart. The physical ports are thus bound to the netmap device driver. Then we simply bind physical ports to a VALE instance (in this case \emph{vale0}) using the {\tt vale-ctl} command:
\begin{lstlisting}
vale-ctl -a vale0:p1
vale-ctl -a vale0:p2
\end{lstlisting}

\subsection*{p2v scenario}
As for p2p, we need to follow switch specific approaches. The only difference is that we have to consider the virtual interface connecting software switches to VNFs. To interact with virtualized environments such as virtual machines or containers, each switch must create a virtual interface. Snabb, VPP, OVS-DPDK, FastClick, and BESS achieve this using the vhost-user protocol. VALE, on the other hand, achieves this using \emph{ptnet}~\cite{maffione2016flexible}. Some configurations are required on the VNF side to implement the p2v workflow. These are described in Sec.~\ref{experiments}. Here we specifically detail the minimal configuration required for each software switch. 
In particular, for BESS we configure a virtual interface ``v1" using the {\tt PMDPort} module by specifying the name and Unix domain socket path. Then physical interface ``inport" is linked to ``v1" to implement p2v workflow:
\begin{lstlisting}[frame=single]
inport::PMDPort(port_id=0, ...)
in0::QueueInc(port=inport, qid=0)

v1::PMDPort(vdev="name,iface=path,...")

in0 -> PortOut(port=v1.name)
\end{lstlisting}

Similarly, for FastClick, t4p4s, and VPP, we create a virtual interface by specifying the name and socket path through the DPDK ``--vdev" option. Note that, by default, t4p4s does not work with virtual interfaces. We thus disabled some offloading features and recompiled the source code to make it compatible with vhost-user. OVS-DPDK accomplishes the same by setting the type of virtual interface to dpdkvhostuser. The created interfaces behave just like physical ones and they can be linked to rendering the intended traffic steering behavior.

Unlike solutions based on DPDK, Snabb implements its own version of vhost-user backend. Consequently, we create a virtual interface ``vi1" leveraging its customized ``vhostuser" module:
\begin{lstlisting}  
config.app(c,"vi1",vhostuser.VhostUser,...)
\end{lstlisting}
As for VALE, we just create a virtual interface using {\tt vale-ctl} and attach it to a VALE instance which relies traffic from the physical interface to the VNF:
\begin{lstlisting}
vale-ctl -n v0
vale-ctl -a vale0:v0
\end{lstlisting}
 

\subsection*{v2v scenario}
To configure software switches realizing v2v workflow, we simply instantiate two virtual interfaces and interconnect them as described in the p2v scenario.

\subsection*{loopback scenario}
For loopback, physical and virtual interfaces are created and interconnected as described for p2p/p2v scenarios. Note that in the loopback scenario, t4p4s relies on the VMs to modify the destination MAC address of each traversing packet according to the flow table.

\end{document}